\documentclass[a4paper,12pt]{scrartcl}
\usepackage[utf8x]{inputenc}
\usepackage[pdftex]{graphicx}
\usepackage{epstopdf}
\usepackage{slashed}

\newcommand{\qv}{q_{v}}
\newcommand{\Fv}{F_{v}}
\newcommand{\Dv}{D_{v}}
\newcommand{\gv}{g_{v}}
\newcommand{\gammav}{\gamma_{v}}
\newcommand{\piv}{\pi_{v}}
\newcommand{\rhov}{\rho_{v}}

\newcommand{\qvbar}{\bar q_{v}}
\newcommand{\Fvbar}{\bar F_{v}}

\newcommand{\pT}{ \slashed p_{\perp} }
\newcommand{\eT}{e_{\perp}}

\title{A first study of Hidden Valley models at the LHC}
\author{Morgan Svensson Seth}

\begin{document}

\begin{titlepage}
\hspace{130 mm} \makebox{LU TP 11-21}

\hspace{130 mm} \makebox{June 2011}
\vspace{5 mm}

\centering {\large \bf Bachelor Thesis}
\vspace{13 mm}

{\centering \Huge \bf A first study of Hidden Valley models at the LHC}
\vspace{15 mm}

{\centering \large Morgan Svensson Seth }
\vspace{7 mm}

Theoretical High Energy Physics 	\linebreak
Department of Astronomy and Theoretical Physics 	\linebreak
Lund University 	\linebreak
S\"olvegatan 14 A 	\linebreak
SE-223 62 Lund	 \linebreak 

\vspace{4 mm}

{\large Supervised by: Torbj\"orn Sj\"ostrand}

\vspace{30 mm}

\begin{abstract}
\begin{center} \makebox{ \bf Abstract } \end{center}
{\flushleft New} stable particles with fairly low masses could exist if the coupling to the Standard Model is weak, and with suitable parameters
they might be possible to produce at the LHC.
Here we study a selection of models with the new particles being charged under a new gauge group, either $ U(1) $ or $ SU(N). $
In the Abelian case there will be radiation of $ \gammav $s, which decay back into the SM. In the non-Abelian case the particles will undergo
hadronization into mesons like states $ \piv/\rhov $ that subsequently decays.   We consider three  different scenarios for
 interaction between the new sector and the
 SM sector and perform simulations using a Hidden Valley model previously implemented in \textsc{pythia}.
In this study we illustrate how one can distinguish the different models and measure different parameters of the models under conditions like those
at the LHC.

\end{abstract}

\end{titlepage}

\newpage
\section{Introduction}
The LHC does not only allow us to continue the search for the Higgs boson but opens up possibilities to find other entirely new
types of particles. One of them is some kind of hidden particle, a particle that interacts very weakly with ordinary matter.
More interesting still is if there are several of them and these new particles interact with each other, creating a whole new
hidden sector \cite{Strassler:2006im,Strassler:2006qa,Han:2007ae,Strassler:2008bv}. This extension with new hidden particles 
can be reasonable from theoretical considerations. In many theories like 
string theory, supersymmetry, grand unification theories etc. one has large symmetry groups, inplying new particles. The Standard Model (SM)
 \begin{math} U(1)_{Y} \times SU(2)_{L} \times SU(3)_{C} \end{math} and its three families of quarks and leptons
 is only a small part of this, and a lot of new interactions between both ordinary and new matter will arise from these symmetry
 groups. These states are usually assumed to be around Grand Unification or the Planck mass, but it is not unreasonable 
to assume that some of them are light, just like in the SM. And  even in the SM there are neutrinos
that are somewhat hidden by only interacting via the massive $ W $ and $ Z $ bosons, so there is no reason why new particles cannot be hidden.
 In addition, if amongst this new sector there is a lowest energy state that is stable, it is a suitable candidate for dark matter. 

In general the possibilities of finding new particles at LHC are firstly if we find particles that previously thought to be 
elementary are in fact composite. If the confining forces are strong enough the composite structures will not be seen unless
sufficient energy is available. Another possibility is if the new particles themselves are too heavy to be produced at previous detectors.
Finally is the model above, that the particles are light but the new particles have no charges in the SM, but couples
 to SM particles through a new coupling that involves a heavy state. These types of theories are known
 under different names such as hidden valley, secluded sector, dark sector etc. and are the ones studied in this paper.
 We will use the term Hidden Valley (HV) to denote them.

Most of the new models like supersymmetry etc. are introduced in order to fix some issues with the SM and as such one introduces
specific gauge groups and particles to deal with these issues. Here on the other hand we will not consider why they are 
introduced from a theoretical perspective, only note that there are theories that allow them. The relevant mass scale in such
theories is not well specified. Since the LHC is the available machine to discover them, it is interesting to study such scenarios
that may give visible effects at the LHC but not at lower energy machines. The models investigated are 
thus picked  with regard to having some visible consequences at the LHC scale. Then the different scenarios are simulated
and the possible signals from different models introduced in a detector like those at the LHC are investigated.
Studies of visible signals from HV scenarios have been presented e.g. in \cite{Carloni:2010tw,Carloni:2011kk}. Since the study in
\cite{Carloni:2011kk} was only for lepton colliders and since the available accelerator in the near future is the LHC, we naturally want to
extend the study for hadron colliders and the LHC in particular.

The paper is structured as follows. In section 2 we begin with an overview of the HV scenarios we study, the different choice of gauge groups
and means of production. Section 3 is a short introduction about the  means of detection at a hadron collider. 
The study begins in section 4 with an overview of parameter choice, followed by the result of the simulations. Finally in section 5
is a summary and conclusions.

\section{Hidden valley scenarios} 

Since we want to investigate if one can find and study the properties of a hypothetical hidden sector at the LHC, we need to consider 
scenarios that produces signals visible at the LHC. Also, in order to study the physics in this new sector,
the hidden particles must decay or radiate back to the SM, since otherwise the only signal is missing energy and transverse momentum. 
This will not gives us much information on what actually goes on in the hidden sector.
 There also has to be some particles that don't decay, otherwise it is not really a hidden sector.

\subsection{Hidden gauge groups}
 We assume that the Hidden Valley consists of a new gauge group G, here assumed to be a $ U(1) $ or $ SU(N) $ group.
There also is a set of fundamental particle(s) $ \qv $ with charges only under the new gauge group. The $ \qv $ is considered
to be a stable particle and it could be either a fermion or boson, but to be consistent with other spin choices we make in
this study it has to be spin $ 1/2 $. In the case of an $ U(1) $ group we will have a photon-like gauge boson, called the $ \gamma_{v} $.
The $ \qv $ will radiate $ \gammav $ as $ \qv \rightarrow \qv \gammav $.
This gauge group is assumed to be broken since otherwise neither  $ \qv $ nor the radiation will decay,
so there will be no visible decays from the HV.
With the broken gauge group we can assume some kind of mixing between $ \gammav $  and the ordinary
photon, so the $ \gammav $ will decay with a lifetime dependent on the mixing angle, into fermion pairs like
 an off-shell photon with the mass of  $ \gammav $.

In the case of a non-Abelian gauge group the gauge bosons, now called $ \gv $, 
will self-interact and if they are massless the interaction
 strength will not fall off at larger distances. This leads to confinement just like in QCD.
 There will also be radiation as $ \qv \rightarrow \qv \gv$ and $ \gv \rightarrow \gv \gv $.
The confinement leads to hadronization of the  
$ \qv $ and $ \gv $ into objects similar to mesons and hadrons (and also possibly some glueball-like state). Since the majority of 
produced hadrons in QCD is the lightest mesons $ \pi $ and $ \rho $, only their 
Hidden Valley counterpart $ \pi_{v} $ and $\rho_{v} $ are included. These particles
are simply assumed to have twice the $ \qv $ mass. 
The ratio of $ \pi_{v} $ to $\rho_{v} $ produced is set to $ 1:3 $ simply from spin counting.
 Now the pair of $ \qv $ and $ \bar\qv $ will be kept confined within 
these hadrons and thus can annihilate by whatever
 method introduced to create them. Since their mass are identical there will be no decay from $ \rhov $ to $ \piv $. Similar to the 
SM there will be an extra factor $ m_f^2 $ for  $ \piv $ decaying into a pair of fermions $f$ due to a change in helicity. To keep some 
hidden particles stable one must assume several flavors of $ \qv $.
 Now the flavor diagonal mesons will decay but the others will be stable. 
Since any flavor of $ \qv $ is created when the string breaks at hadronization, with $ n $ flavors
 roughly $ 1/n $ of the mesons will decay.

Thus the models will generate invisible particles, but also some signals of decay back to the standard model 
which, allows for measurements and a possibility to determine the hidden valley physics.

\subsection{Production}
For production, the simplest way to imagine interactions between the hidden and normal 
sectors is to introduce a heavy boson 
$ Z' $ with coupling both to standard model and Hidden Valley particles. Then the
$ \qv \qvbar $ pair is produced from the SM sector via a $ q \bar q \rightarrow Z' \rightarrow \qv \bar \qv $.
 The $ Z' $ will be assumed to have a mass of around 1 TeV in order to have remained hidden at previous
 accelerators, but to be light enough to be produced at the LHC.

Another way is to introduce a particle $ \Fv $ with charge under both SM and the
new gauge group. The $ \Fv $, has to be a boson in order to be consistent with $ \qv $ being 
spin $ 1/2 $ in decays. Then it can be pair-produced through an ordinary 
QCD process such as $ q \bar q / gg \rightarrow \Fv \Fvbar $ or for a fermion $ f $ as
 $ f \bar f \rightarrow \gamma / Z^{0} \rightarrow \Fv \Fvbar $. Since these processes occur at SM rates
 one must assume that the $ \Fv $ has a mass of several hundreds of GeV to ensure
that it has not been observed at previous detectors. Since these particles are charged under the hidden gauge group they will radiate
$ \gammav $ or $ \gv $ and also standard model radiation based on their
 SM charge. If kinematically possible the $ \Fv $
state will decay to one SM fermion and a $ \qv $  $ \Fv \rightarrow f\qv $. In order to preserve quantum numbers the 
decay has to be flavor diagonal, so one must introduce one $ \Fv $ for each standard model fermion.
Naming of these is with uppercase letter for the respective particle, such as $ \Dv \rightarrow d \qv $
and $ E_{v} \rightarrow e^{-} \qv $. Production at a hadron collider will primarily be through the strong interaction,
so we use the $ \Dv $ as a typical produced particle. All different $ \Fv $ are still included for
 the decay of the hadronic states.

Finally one can use the $ \gamma/\gammav $ mixing  to mix an off shell-photon into a valley photon and
thus pair-produce $ \qv $. The nature and origin of the mixing is unspecified, but the key parameter is a
mixing angle between the two states. Although the production mechanism is through $ \gammav $ we will
still also consider the alternative with a $ SU(N) $ gauge group in the hidden valley. 

\section{Detection at a hadron collider}
The difficulties of detecting particles at a hadron collider arises since the colliding protons are composite objects,
and the desired interaction is only one of possibly many occurring between the partons of the protons. This means there is a large
background, and since quarks hadronize to a large set of light mesons and baryons the outgoing particles will, apart from leptons,
 not be easily distinguishable from the background. 

Since a parton only carries a fraction of the total proton
  momentum, the interesting subcollision will often
have some initial momenta in the beam $ (z) $ direction. The beam remnants escape detection through the beam pipe, so the total $ z $
momenta cannot be measured. Only the transverse part of the momentum, $ p_{\perp} $, and correspondingly
 $ \eT = \sqrt{m^{2}+p_{\perp}^{2}} $ is in many cases used.

The actual interaction of interest at a hadron collider
 will, due to the initial momenta in the z direction, not be spherically symmetric. However, there should be an (approximate) symmetry under Lorentz boosts
 in the $ z $ direction of the hard colliding subsystem.
As such a good parametrization is using $ (\eta, \phi, \eT) $ where $ \eta $ is
the pseudorapidity $ \eta=\frac{1}{2}\ln{\frac{|p|-  p_{z}}{|p|-p_{z}}} $. It is an approximation, in the massless limit, to 
ordinary rapidity, which is additive under Lorentz boosts.
As such many distributions will be fairly even in the pseudorapidity. To find high $ p_{\perp} $ quarks or gluons one can search for their 
hadron jets by looking inside a circle in the
$ \phi,\eta $ plane. If sufficient amount of $ \eT $ is present one considers this as a jet. With proper radius compared to the $ \eT $
one can ensure that for some given decay the products, if possessing enough momenta, also will sit inside
the circle. This means if the original particle does not possess enough momentum it will be missed, but on the other hand lowering the $ \eT $
limit means risk of catching several background jets.
For jet finding we use \textsc{Pythia}  built-in
jet finder CellJet. It has some flaws, like if two jets overlap all of overlap goes to the first found jet, but it is sufficient 
to show the main principles of HV jet distributions.

Sphericity is a measure of how round an object is, and in particle physics is used to evaluate how evenly the momenta of
detected particles is distributed.  One defines a sphericity tensor as
\begin{equation} S^{ij}=\frac{\sum_{m}{p_{m}^{i}p_{m}^{j}}}{\sum_m{ |\bar p|^{2}}},
\end{equation}
where m runs over all particles of the event. Then the eigenvectors of this matrix defines rotational axes and the
eigenvalues $\lambda $ are measures of the length of the object in said direction. Then if $\lambda_{1} $ is the largest eigenvector,
 the sphericity is defined as
\begin{equation}  \frac{3}{2}\frac{\lambda_{2} + \lambda_{3}}{\sum_{k}\lambda_k}. \end{equation} 
The factor 3/2 means that the sphericity lies between unity, for a sphere, and zero, for a linelike object.
The original definition above will lead to a quadratic dependence on momenta, and whether a particle decays or not will influence the result.
One therefore often uses a linearized version of the sphericity tensor
\begin{equation} S^{ij}=\frac{\sum_{m}({p_{m}^{i}p_{m}^{j}}/|\bar p_{m}|)}{\sum_m{ |\bar p|}}. \end{equation}
An event at a hadron collider is rather cylindrical due to the beam remnants, so only the two dimensional transverse part is of interest.
To keep the range from 0 to 1 the expression now becomes $ 2\lambda_{2}/(\lambda_{1}+\lambda_{2}) $.

\section{Analysis of the hidden valley scenarios}
To study the different scenarios we will use the \textsc{Pythia} \cite{Sjostrand:2007gs} 
event generator. \textsc{Pythia} uses Monte Carlo methods to simulate the 
entire collision process, beginning from selecting partons from parton density functions and
 evaluating hard-process cross sections, on to initial and final radiation, string fragmentation and hadronization, beam remnants 
and decays. In total one obtains an end result similar to what one can observe in an actual detector but with the additional
benefit of knowing how one got there. In the studies here we will use statistics from 10000 events in which the
 respective HV process in
question actually did occur. The different scenarios will be denoted with Z for the $ Z' $ mediated,
Fv for the $ \Fv $ mediated and KM for the mixing of the $ \gammav $ with the photon. In addition we will affix 
these with A for Abelian and NA for a non-Abelian gauge group.

\subsection{Parameters}
In total we have six models with several new particles and interactions, so there are many parameters. Some of them will
have constraints from measurements in previous detectors. Since we are studying possible dark matter candidates
we also have constraints from cosmological observations. Neither of these constraints will be explored here. Since the methods of measurement will be fairly similar for somewhat 
different parameters, we simply picked reasonable values. We will only consider when LHC is up 
at full energy with 14 TeV collisions.

 For the model parameters first one has to consider the production
cross sections. This will depend only on the $ \gammav $ mass and the mixing angle in the KM scenario.
For the other scenarios it depends on the masses of $ \Fv $ and $ Z' $ respectively, and their respective couplings to SM and HV. For the kinematics
one also have to determine the masses of all involved particles.

For the non-Abelian scenario the hidden gauge group is picked to be $SU(3)$. The HV gauge group coupling strength enters to determine the amount of
 radiation, along with the masses. One also need the number of $ \qv $
flavors to determine the fraction of decaying flavor diagonal mesons,which is put to three for non-Abelian scenarios. The lifetime 
of the $ \gammav $ depends on the mixing angle (for $ \piv/\rhov $ it is the mass of $ Z'/\Fv $) and if high there might be displaced vertices.
Since displaced vertices will only make detection easier we assume there is none.

 Here we will not attempt to make a 
realistic study of issues with background of other events, but simply stay with studies of the HV signal.
Assuming sufficient data for statistics we can ignore the actual production probability and as such the mixing angle. 
 There is still many parameters left so its not suitable for a any real exploration of the
parameter space. These additional parameters are kept at their default values, unless otherwise noted. Changing variables is then only used 
to highlight some important dependence. The default values are as follows; also see Table \ref{parameters}.
\begin{table}
  
\centering
\begin{tabular}[t]{|c|c|c|c|c|c|c|c|c| }
\hline
Model & $ \alpha_{HV} $ & $ N_{colour} $ & $ N_{flavor} $ & $ m_{\qv} $ & $ m_{\gammav} $ & $ m_{\rhov/\piv} $  & $ m_{\Fv} $ & $ m_{z'} $ \\
\hline
ZA & $ 0.1 $ & $ 1 $ & $ 1 $ & $ 100 $ & $ 10 $ & $ - $  & $ - $ & $ 1000 $ \\
\hline 
ZNA & $ 0.1 $ & $ 3 $ & $ 3 $ & $ 5 $ & $ - $ & $ 10 $  & $ - $ & $ 1000 $ \\
\hline
KM & $ 0.1 $ & $ 1 $ & $ 1 $ & $ 100 $ & $ 10 $ & $ - $  & $ - $ & $ - $ \\
\hline
KMNA & $ 0.1 $ & $ 3 $ & $ 3 $ & $ 5 $ & $ (10)$ & $ 10 $  & $ - $ & $ - $ \\
\hline 
FvA & $ 0.1 $ & $ 1 $ & $ 1 $ & $ 100 $ & $ 10 $ & $ - $  & $ 400 $ & $ - $ \\
\hline  
FvNA & $ 0.1 $ & $ 3 $ & $ 3 $ & $ 5 $ & $ - $ & $ 10 $  & $ 400 $ & $ - $ \\
\hline
\end{tabular} 
\caption{Tables of the different parameters for the different scenarios. All masses are in GeV}
\label{parameters}
\end{table}

 The coupling strength 
$ \alpha_{HV} = 0.1 $, the $ \Fv $ are all assumed to have the same mass of $ 400 $ GeV  and $ M_{Z'} = 1 $ TeV. Due to the ease of detection
through lepton pairs we put $ M_{\rhov} = M_{\piv} = M_{\gammav} = 10 $ GeV to get  one of the easiest observable to the same value
for all models when we try to distinguish them. With the mesons at twice the $ \qv $ mass this means that all flavors of 
$ \qv $ have the same mass at 5 GeV in the non-Abelian case. For the Abelian case there are restrictions from dark matter 
experiments such as \cite{Adriani:2008zr,Adriani:2008zq} so then we put $ M_{\qv} = 100 $ GeV.

\subsection{Analyses}

\begin{figure}
\centering
\includegraphics[width=135 mm, angle = 0]{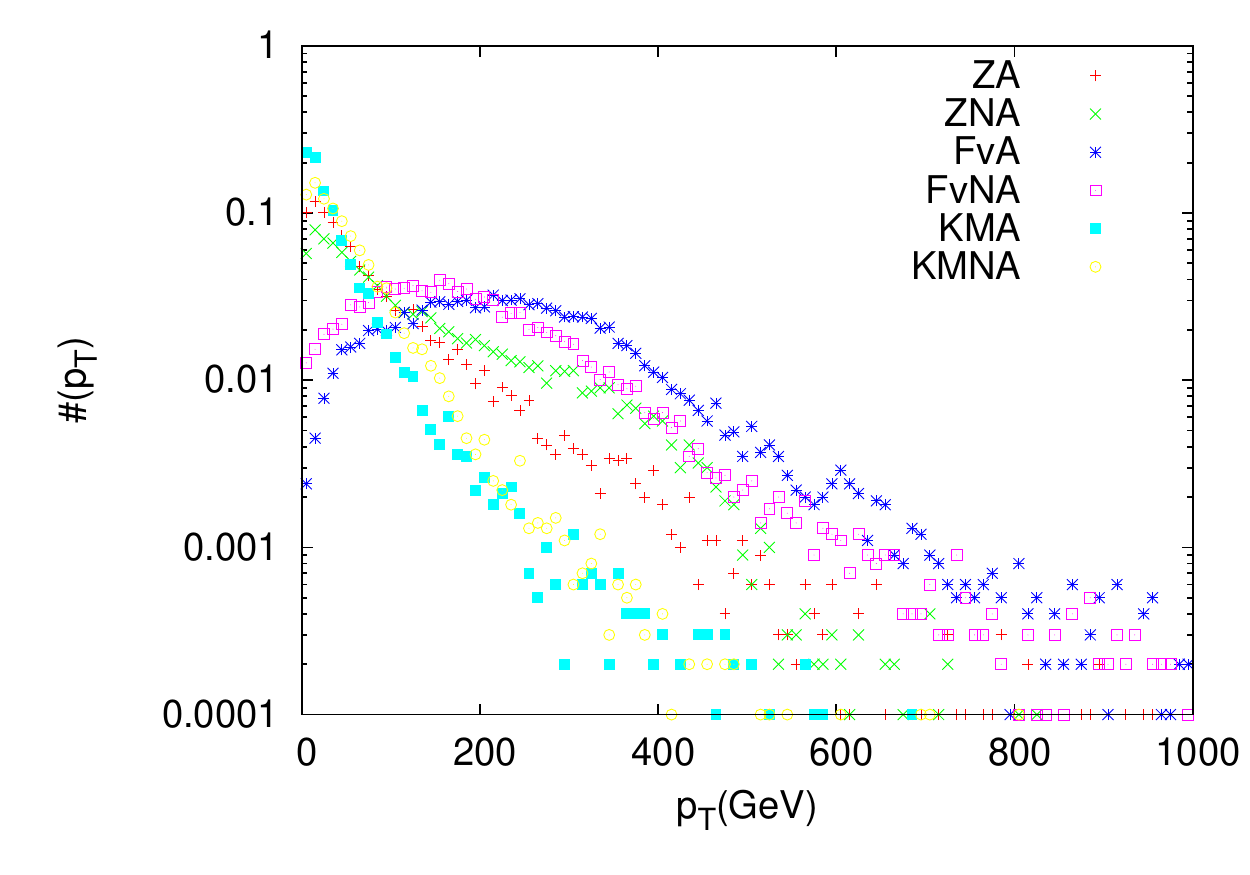}
\caption{ $ \pT $ for the different scenarios using default values for parameters.}
\label{pT}
\end{figure}

Since there will always be Standard Model events that outnumber the HV ones, one must first study distributions
that can at least do a reasonable job of separating HV signals from the background.
 Since the events consist partly of hidden particles, namely the $ \qv $ in 
the Abelian case and non-diagonal $ \piv $ and $ \rhov $ in the non-Abelian one, the missing transverse
 momentum serves as an obvious first choice, Fig. \ref{pT}.
The  $ \pT $ is in general much larger than in SM events and as such arise mostly from the HV effects, although neutrinos are present as well. 
Due to conservation of momenta the $ \qv $ pair from the $ Z'/\gammav $ decay will be back-to-back
 in their rest frame, so only differences in the $ \qv \rightarrow SM $ decays and the original
momentum of $ Z'/\gammav $  give rise to missing momentum from the HV.
Inherently the $ Z' $ and  $ \gammav $ mediated scenarios are
similar but the $ Z' $ mediated has a bit higher $ \pT $. This comes from the $ Z' $ being mainly produced on shell 
at 1 TeV, which leads to energetic $ \qv $, while the $ \gammav $ has to be off shell to even reach the 
200 GeV needed to produce a $ \qv $ pair. The Fv model gives rise
to a much higher $ \pT $ since the $ \Fv \rightarrow f\qv $ decays can happen in a similar direction for both $ \Fv s $.
This is also seen in that the others increase towards no $ \pT $ due to no emissions at all, while the Fv events
rarely line up perfectly, so there is an decrease in number of events when $ \pT \rightarrow 0 $.
There is also some differences between the Abelian and non-Abelian setup, especially
visible in the Z scenarios. It arises since the probability of emitting a certain total amount of energy as $ \gammav $ from the $ \qv $ 
will be exponentially distributed, as for ordinary bremsstrahlung. As such the total $ \pT $ will be approximately distributed exponentially.
  For the non-Abelian the total amount of radiation will depend on the 
number of diagonal mesons at hadronization, and the $ \pT $ spectrum will not be exponentially distributed.

\begin{figure}
\centering
\includegraphics[width=78 mm, angle = 0]{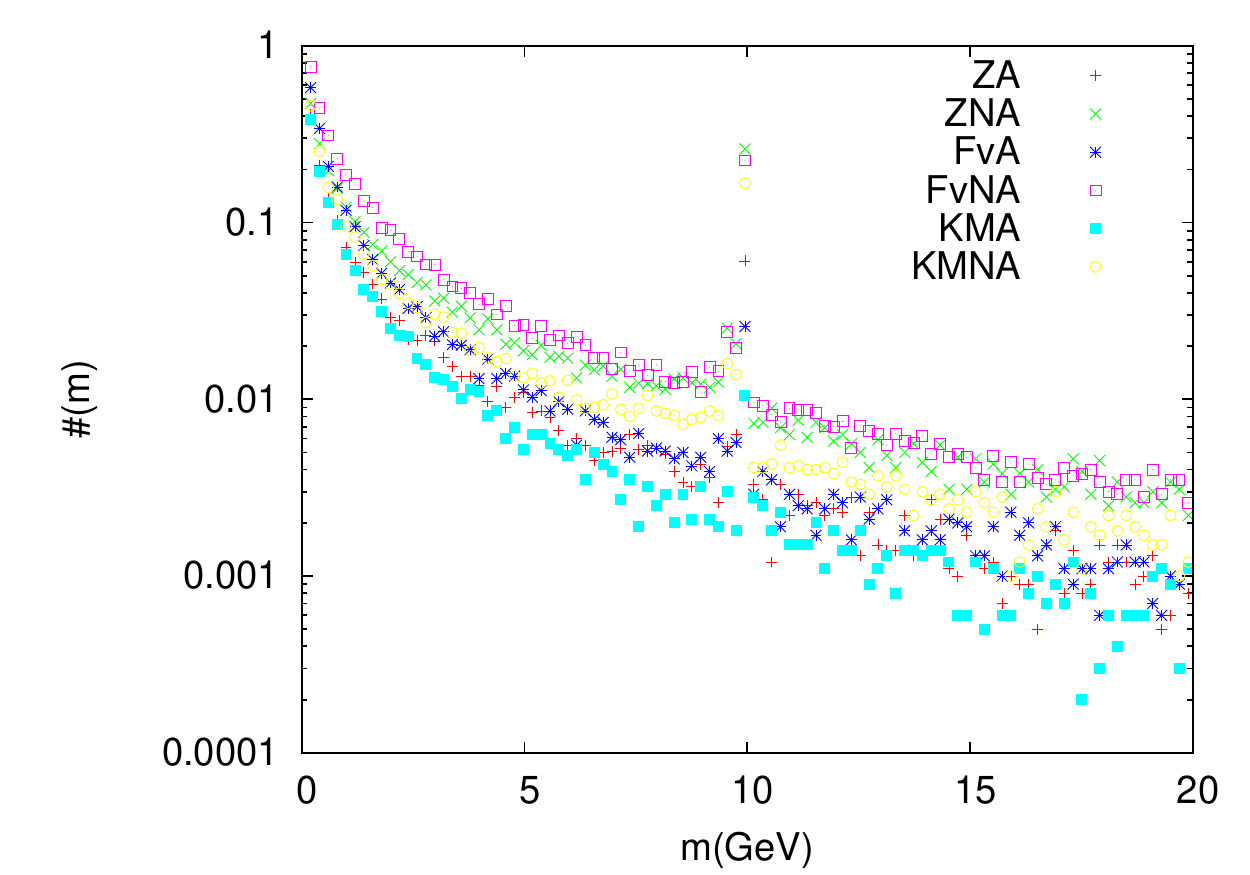}
\includegraphics[width=78 mm, angle = 0]{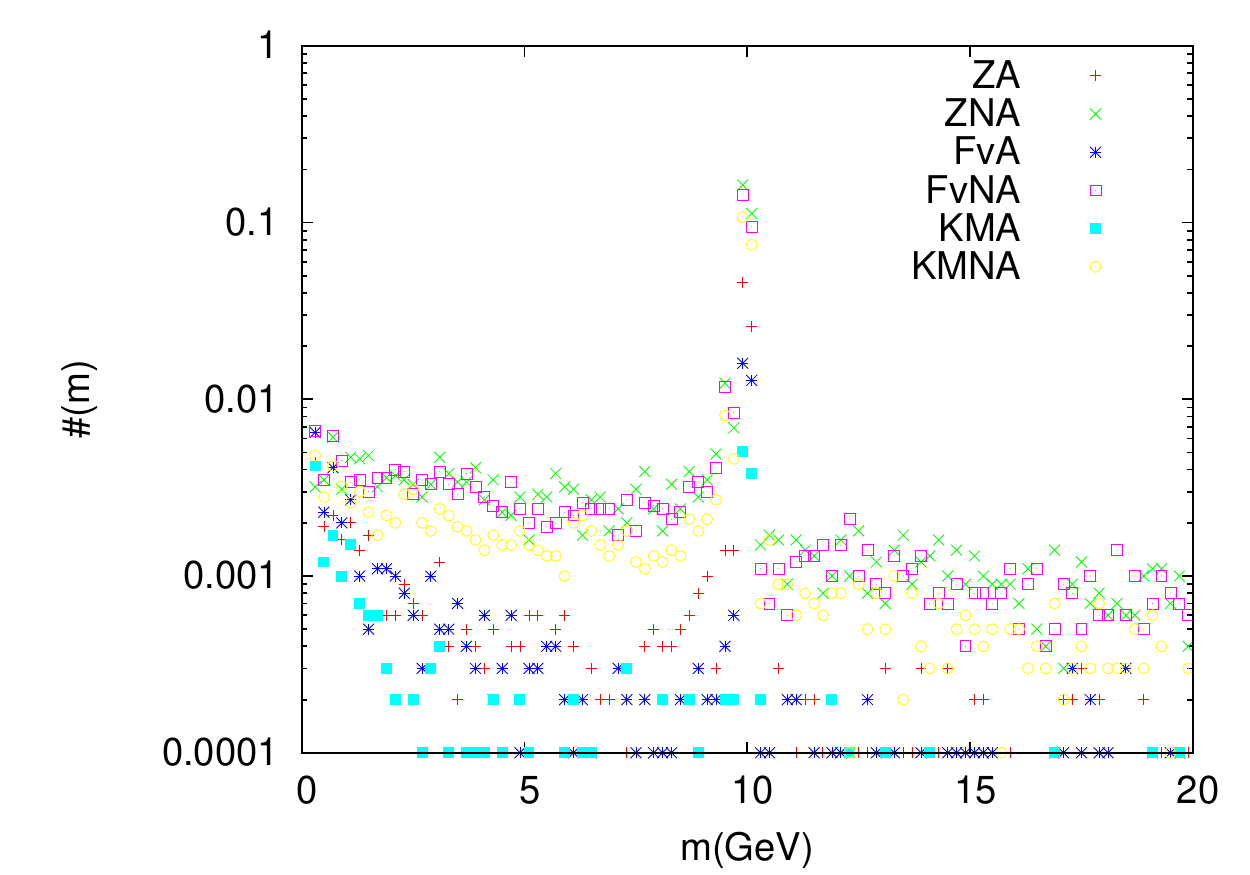}
\caption{The invariant mass of lepton pairs per event. Electrons to the right and muons to the left. Note the spike at 10 GeV, 
the mass of $ \gammav $ or $ \piv/ \rhov $}
\label{lepton}
\end{figure}

Still SM events will obviously in rare cases experience higher missing momentum, especially in weak processes.
Then, in order to detect HV models with 
lower missing momentum, just the  $ \pT $ might not be enough. Another good trigger is the invariant mass of lepton pairs.
 Both the $ \gammav $ and the $\piv/\rhov $ can decay to lepton pairs. Massive such pairs in the SM such as $ J/\Psi, \Upsilon, Z $ etc. 
are reasonably rare and well understood. Also, since there are no strong
interactions involved, the leptons are easy to detect.
The invariant mass of lepton pairs should have
a spike near the mother particle mass, as shown in fig. \ref{lepton}. (The increase of electrons close to zero 
is due to the Dalitz decay $ \pi^{0} \rightarrow \gamma e^{+}e^{-} $
of ordinary pions. This is visible even up to a several GeV due to paired leptons from different pions.)
Detection may be problematic if the mass spike is near a SM one, but in conjunction with high  $ \pT $ at least the mass of
the decaying HV particle can be determined. In addition the presence of both high $ \pT $ and lepton pairs with proper invariant
mass can be a good way to single out the HV events. For the non-Abelian case there might be several complications. The 
mass of $ \piv $ and $ \rhov $ does not need to be the same, the different flavors of $ \qv $ can have different mass 
giving a lot of different lepton signals. Also the $ \piv $ and $ \rhov $ is only the most common of many possible hadrons.

\begin{figure}
\centering
\includegraphics[width=78 mm, angle = 0]{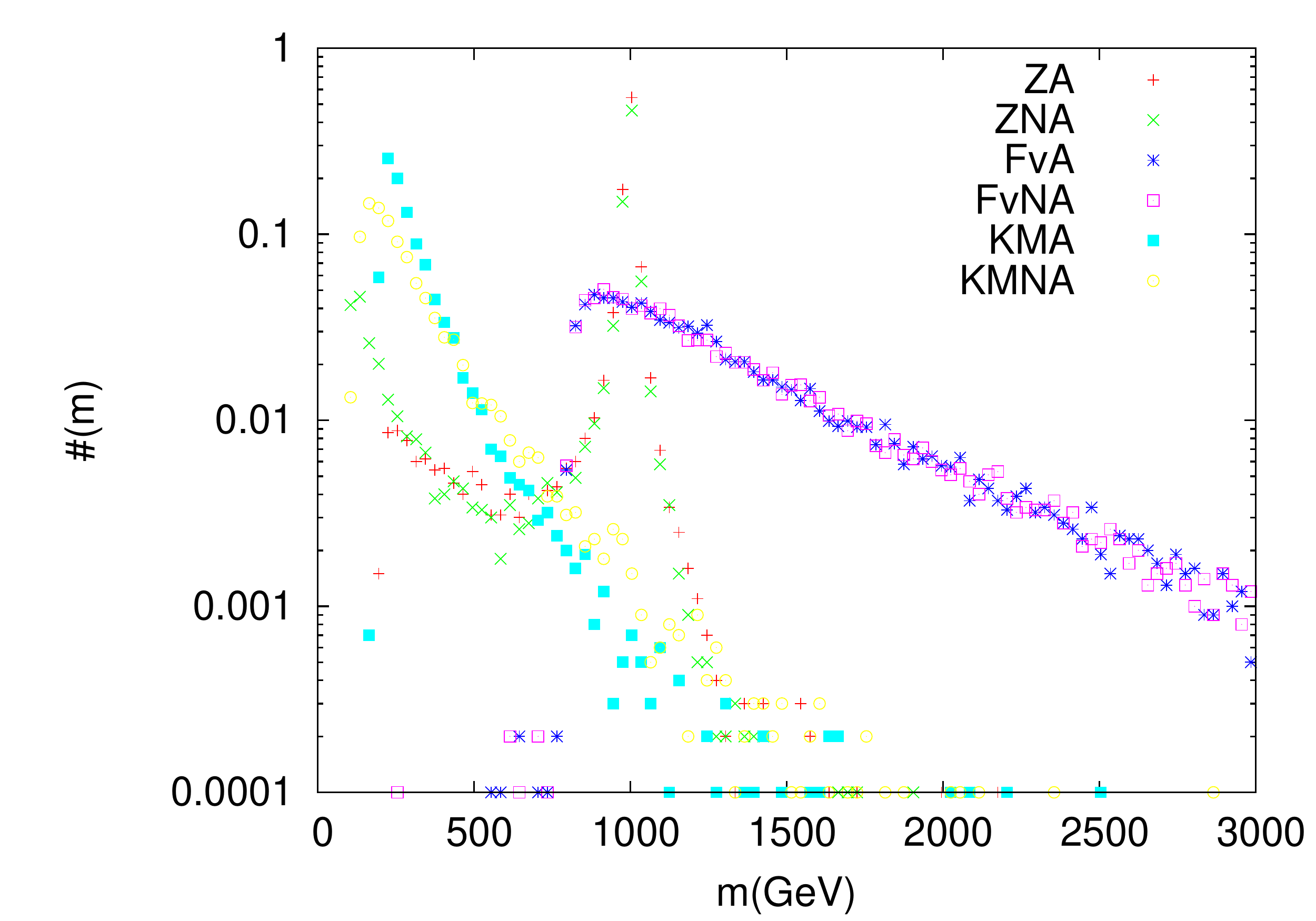}
\includegraphics[width=78 mm, angle = 0]{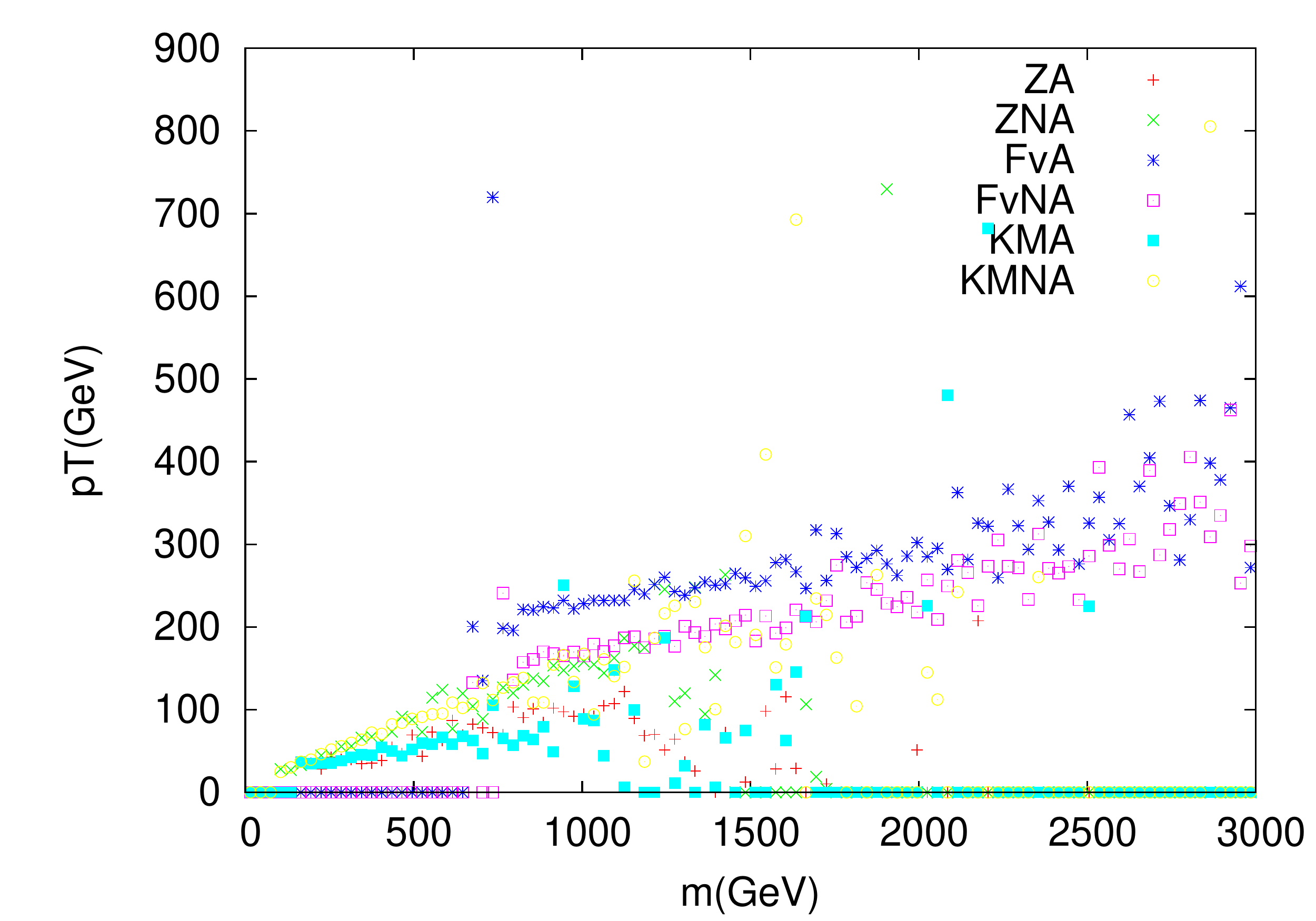}

\caption{To the left the invariant mass of the produced particles, $ \qv $ except for the 
Fv scenarios where it is the $ \Dv $. To the right is missing momentum of the event as a function of the invariant mass.
The momenta for nearby invariant masses  ($\pm 15 $ GeV)  are averaged over.
 The points off on the high and low end of the mass-$ \pT $ plot is due to few events and thus large statistical errors.
}
\label{pTm}
\end{figure}

Although the missing momentum arises from different amounts of decay back to the SM in different directions the amount
should, on average, still depend on the initial mass. A larger mass will give rise to larger energies of the hidden 
particles which allows greater disparity in amounts of decay. We plot the invariant mass of the produced particle pair, $ \Fv $
for the Fv mediated scenarios and $ \qv $ for the rest, in Fig. \ref{pTm}. In particular the mass corresponds to the $ Z' $ mass 
 in its scenarios. KMA events 
are only present above 200 GeV since the $ \qv $ is stable so it needs to be produced on shell. The same effect is present 
for the ZA events, although with much less low-energy events the $ \qv $ mass will be harder to measure.(The lower cutoff 
for the ZNA is a cutoff for the Breit-Wigner distributions in \textsc{Pythia}). Similarly in the $ \Fv $
mediated scenarios most events are above the 800 GeV threshold to produce two on shell $ \Fv $, although there are some off
shell events. For the $ Z' $ the invariant mass corresponds directly to its actual mass so here a mass spike is also present.
 The $ \pT $ as a function of mass is also shown. The $ \pT $ dependence on mass is linear for all studied 
scenarios, so one can use $ \pT $  distributions to constrain mass distributions. Due to the wide difference in $ \pT $ from 
single events one will need a large number of events.

The $ Z' $ mass might be easier to measure from simple SM processes as $ q \bar q \rightarrow Z' \rightarrow l^{+}l^{-} $
in the same way as the ordinary Z boson. Still the right mass scale can be obtained from the $ \pT $ spectra, and the existence
of a $ Z' $ at appropriate mass can be used to distinguish the Z from the KM scenarios. Also for the non-Abelian cases a determination 
of the $ \qv $ mass will encounter lots of difficulties, since one needs low-$ \pT $ events, to determine the low end of the mass 
spectra, and these may not be easily distinguished from SM
events.
Otherwise if one measures a narrow peak in the lepton pairs, how to interpret that as a $ \qv $ mass is a complicated but separate issue,
that we will not consider here.

\begin{figure}
\centering
\includegraphics[width=78 mm, angle = 0]{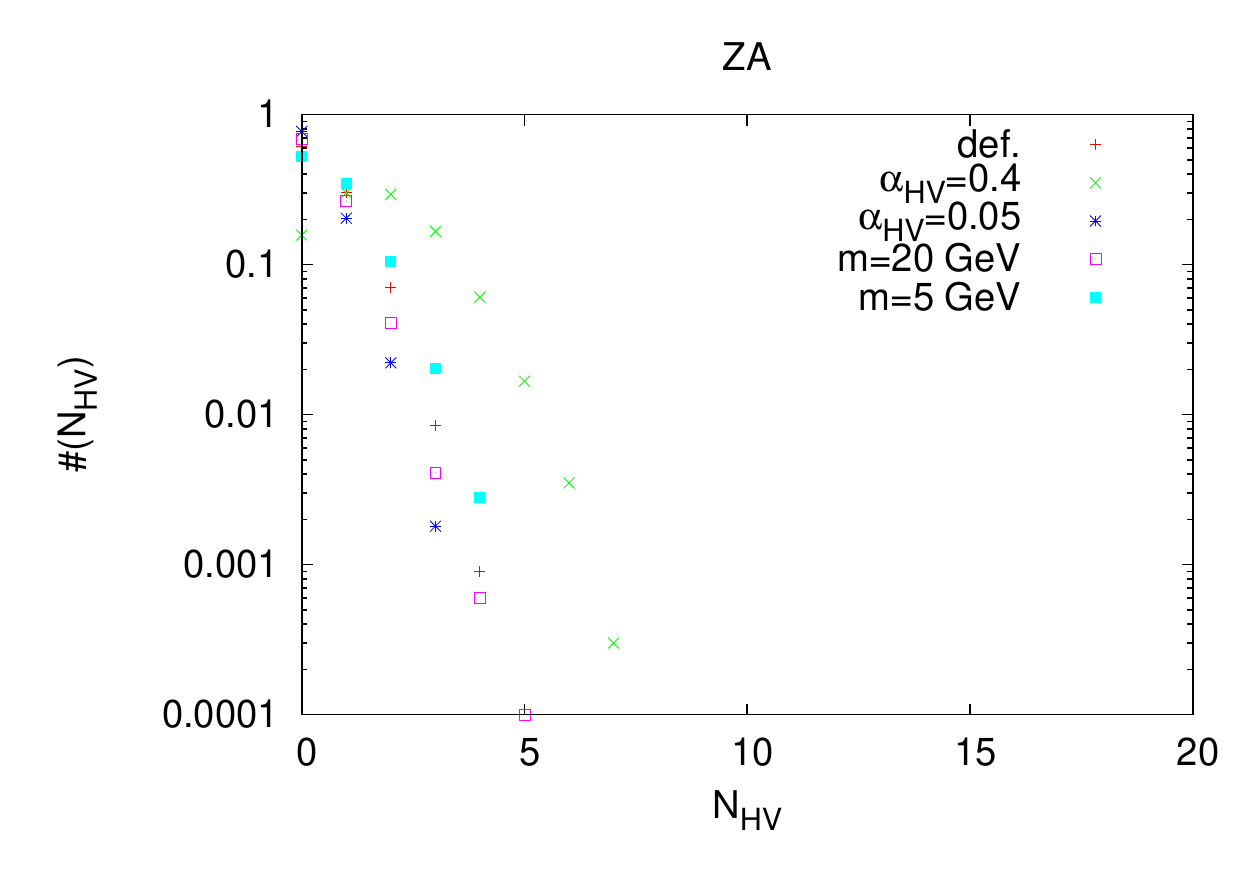}
\includegraphics[width=78 mm, angle = 0]{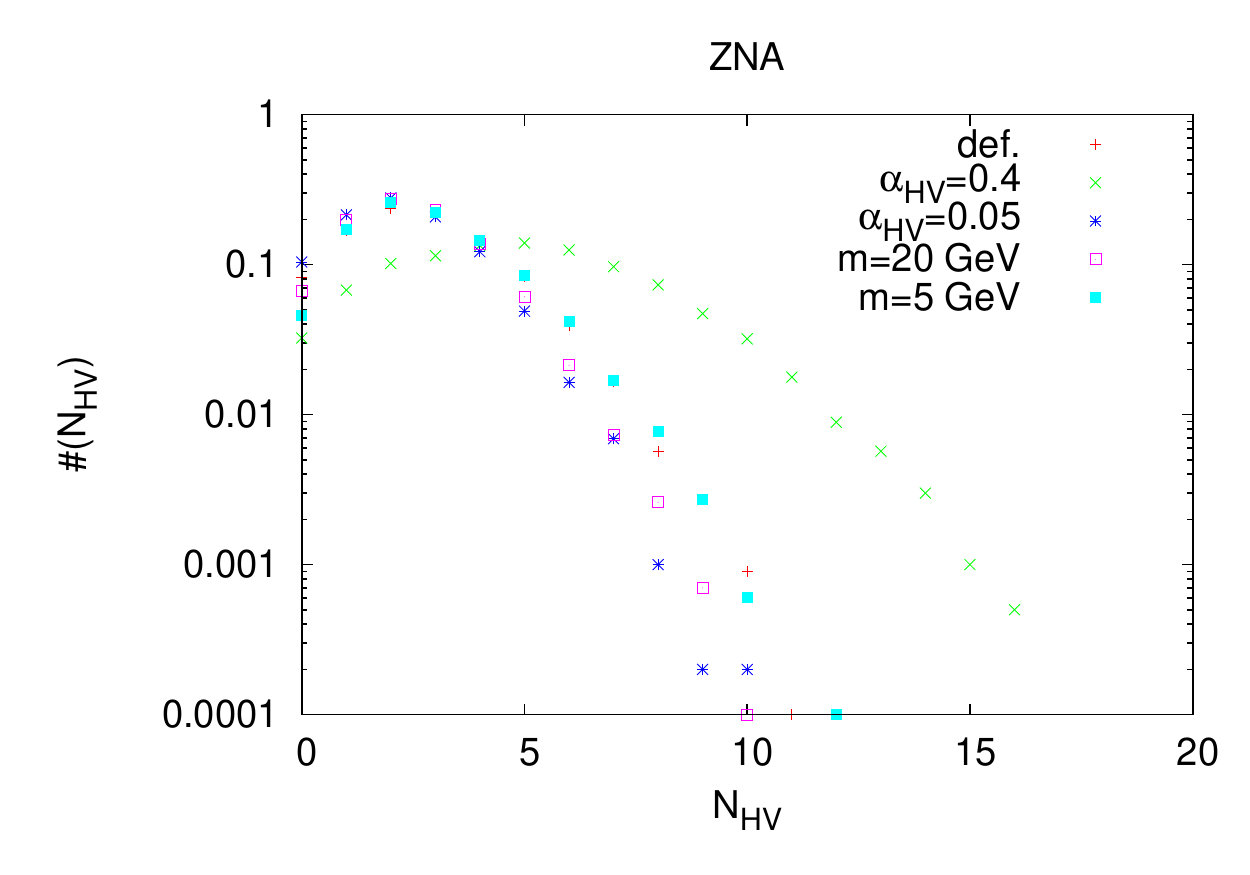}
\includegraphics[width=78 mm, angle = 0]{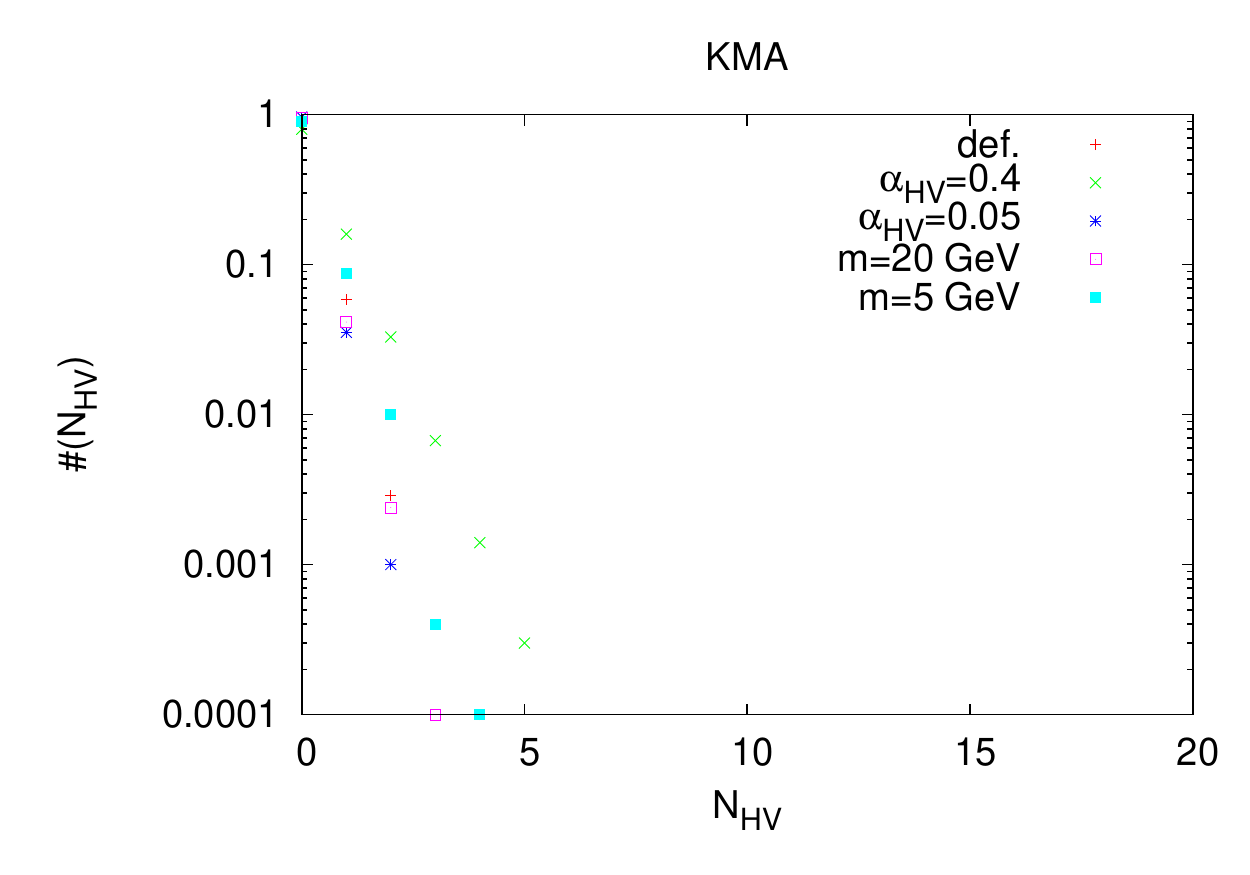}
\includegraphics[width=78 mm, angle = 0]{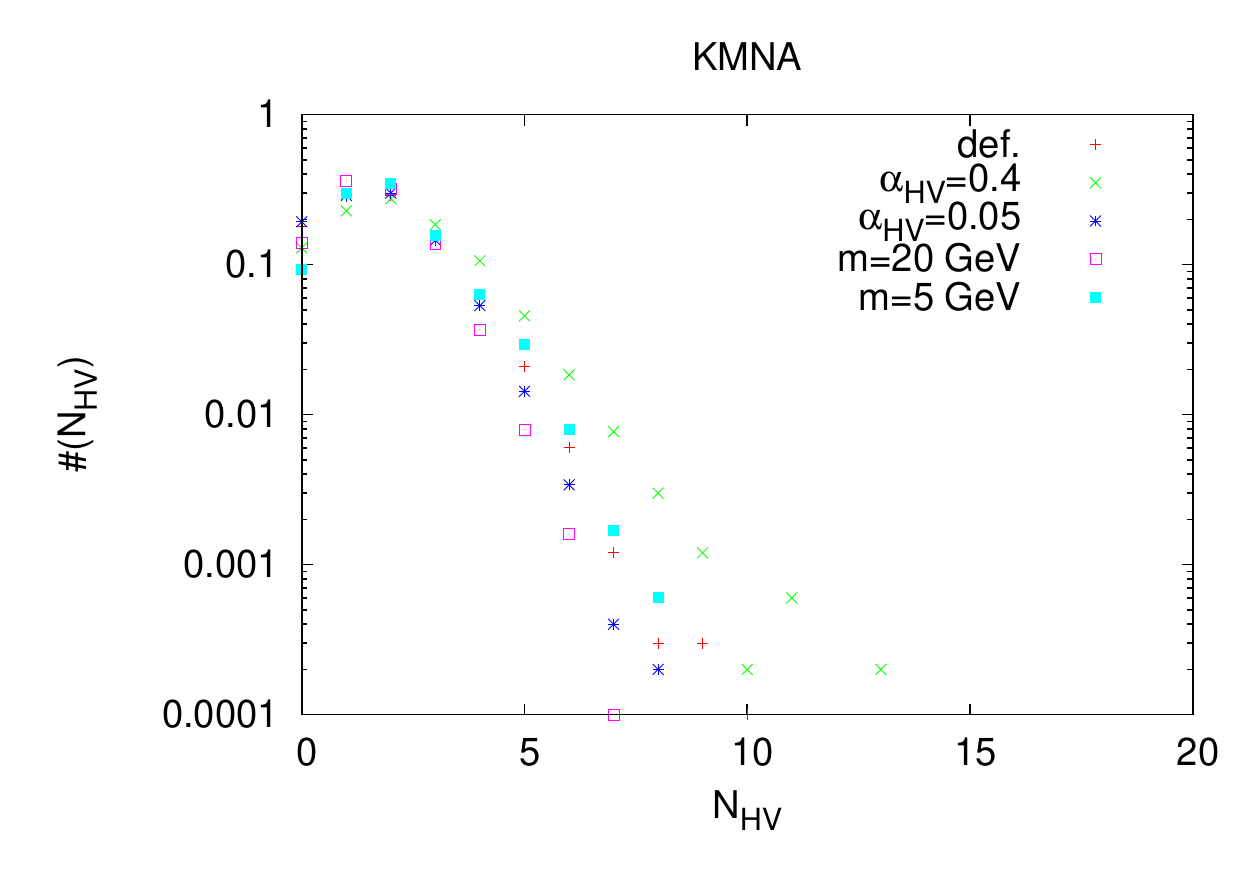}
\includegraphics[width=78 mm, angle = 0]{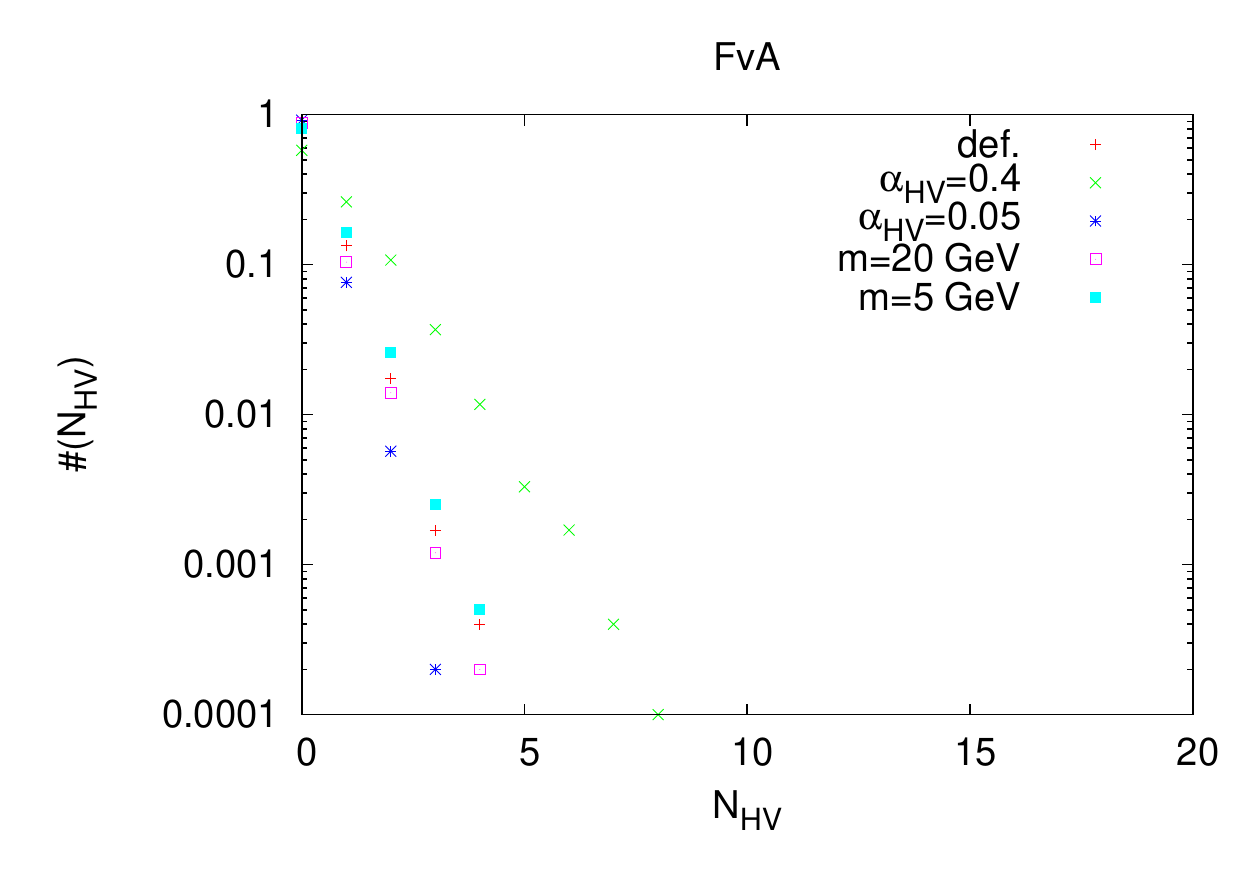}
\includegraphics[width=78 mm, angle = 0]{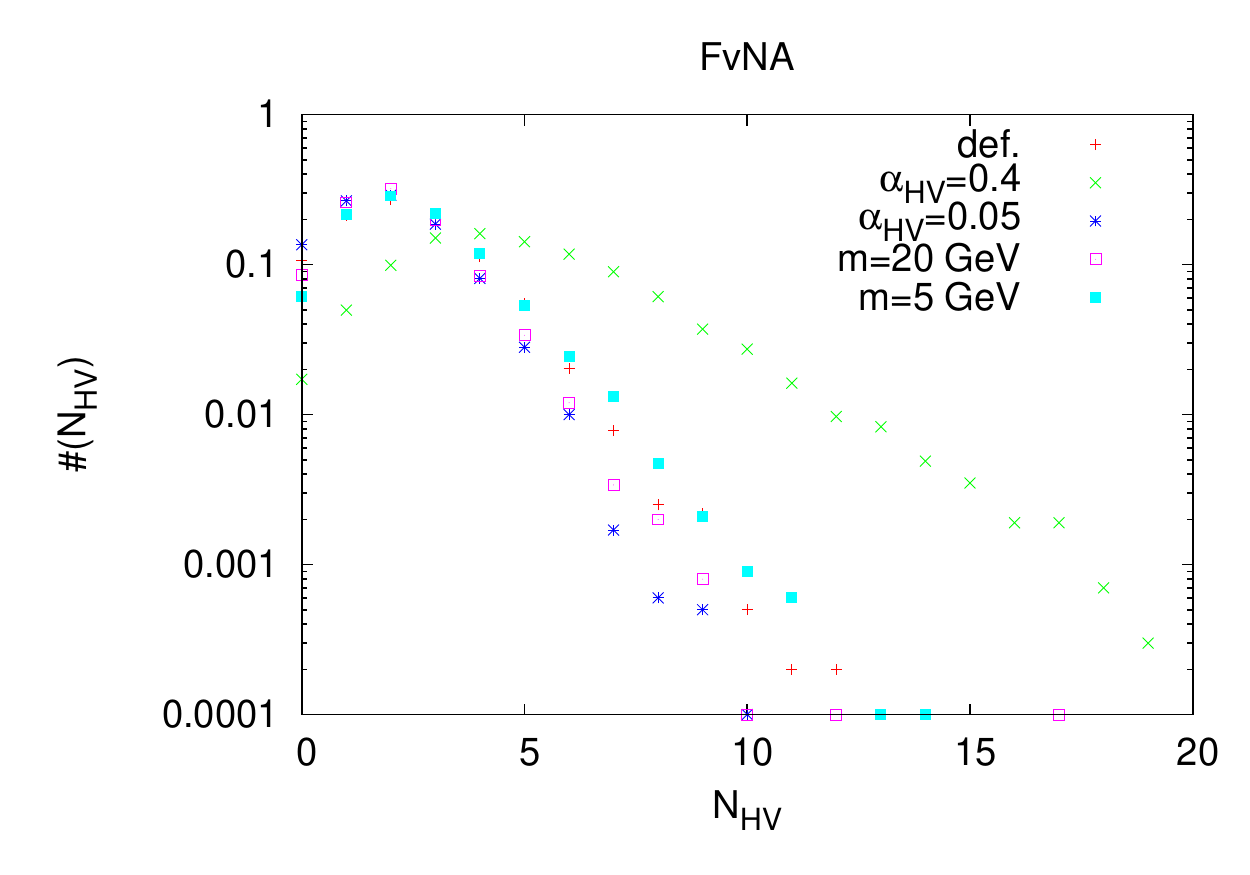}
\caption{The amount of valley particles that decay back in the different scenarios. From top to bottom is Z, KM and Fv, to the left
Abelian and to the right non-Abelian. The numbers will vary with the coupling strength and the $ \gammav /\piv /\rhov $ mass, so they
are varied in the plot. The default values are $ \alpha_{HV}=0.1 $ and $ m_{\gammav /\piv /\rhov}=10 $ GeV}
\label{nRad}
\end{figure}

\begin{figure}
\centering
\includegraphics[width=78 mm, angle = 0]{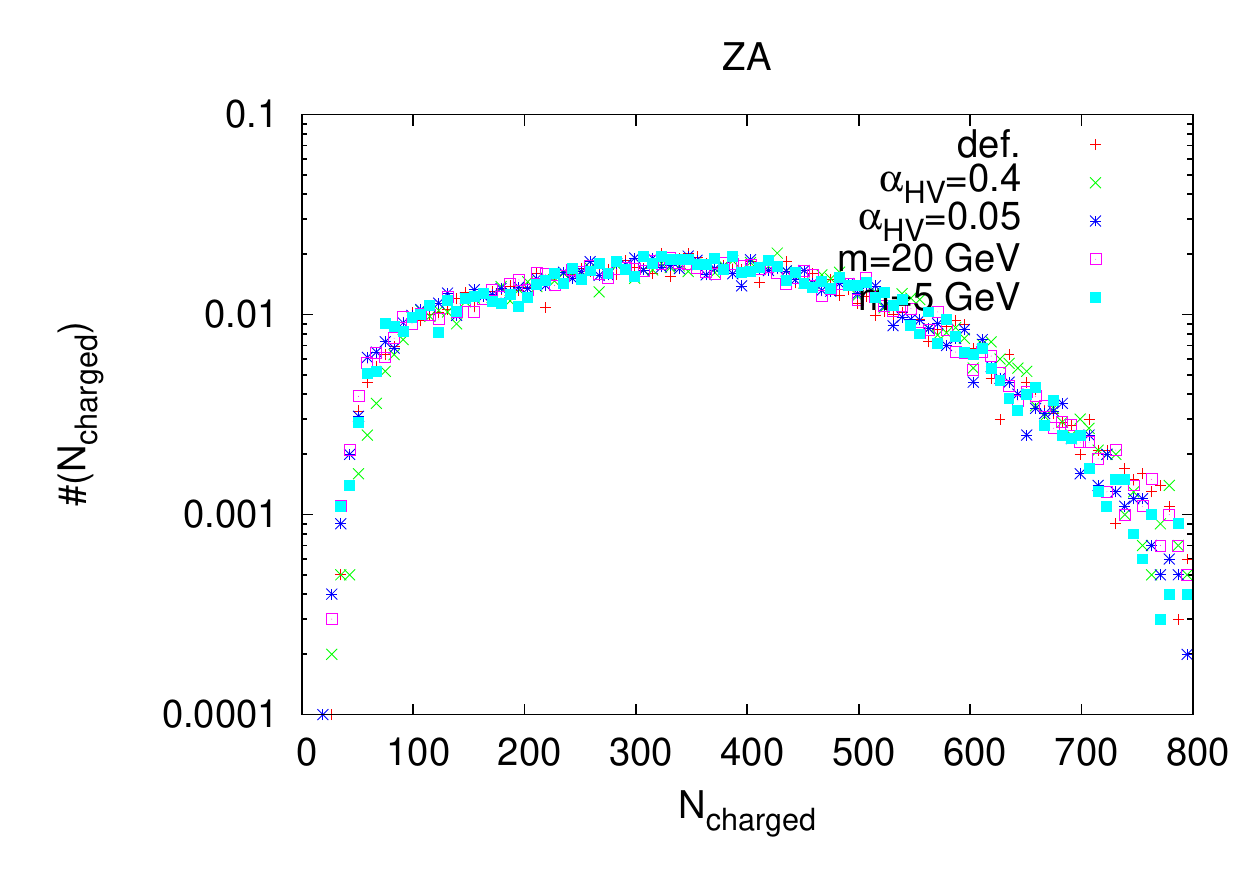}
\includegraphics[width=78 mm, angle = 0]{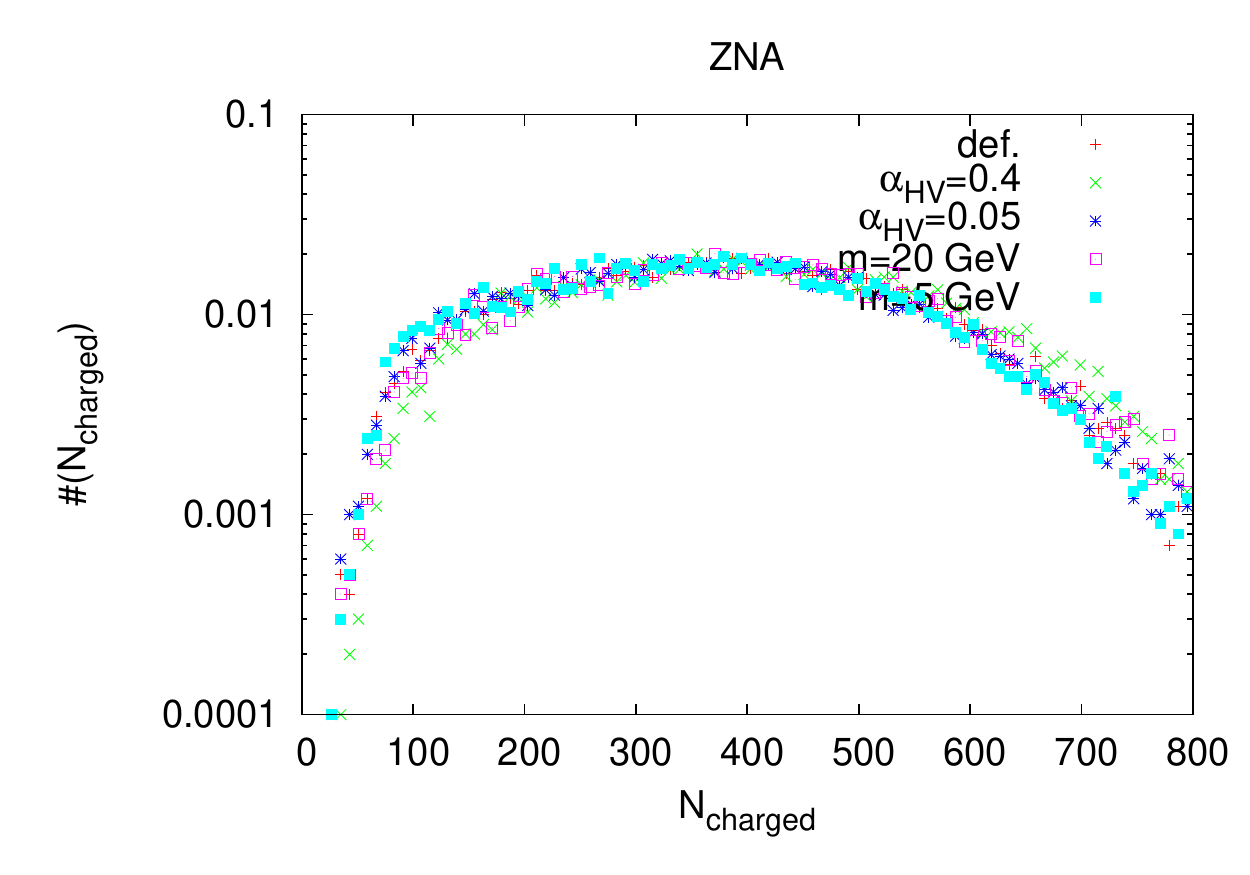}
\caption{Number of charged particles in the ZA (left) and ZNA (right),
 the coupling strength and the $ \gammav /\piv /\rhov $ mass is varied as in fig \ref{nRad}.
The default values are $ \alpha_{HV}=0.1 $ and $ m_{\gammav /\piv /\rhov}=10 $ GeV}
\label{nCharged}
\end{figure}

\begin{figure}

\includegraphics[width=78 mm, angle = 0]{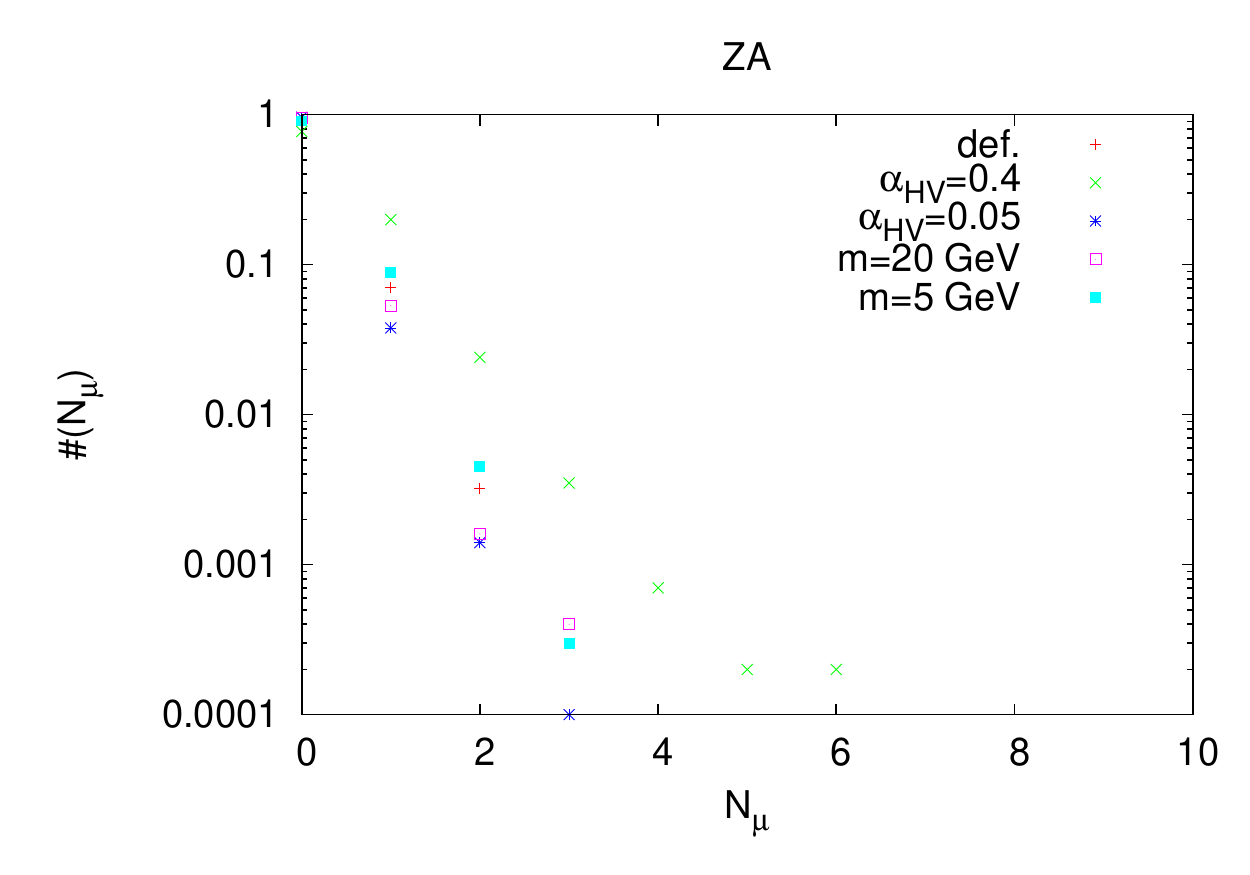}
\includegraphics[width=78 mm, angle = 0]{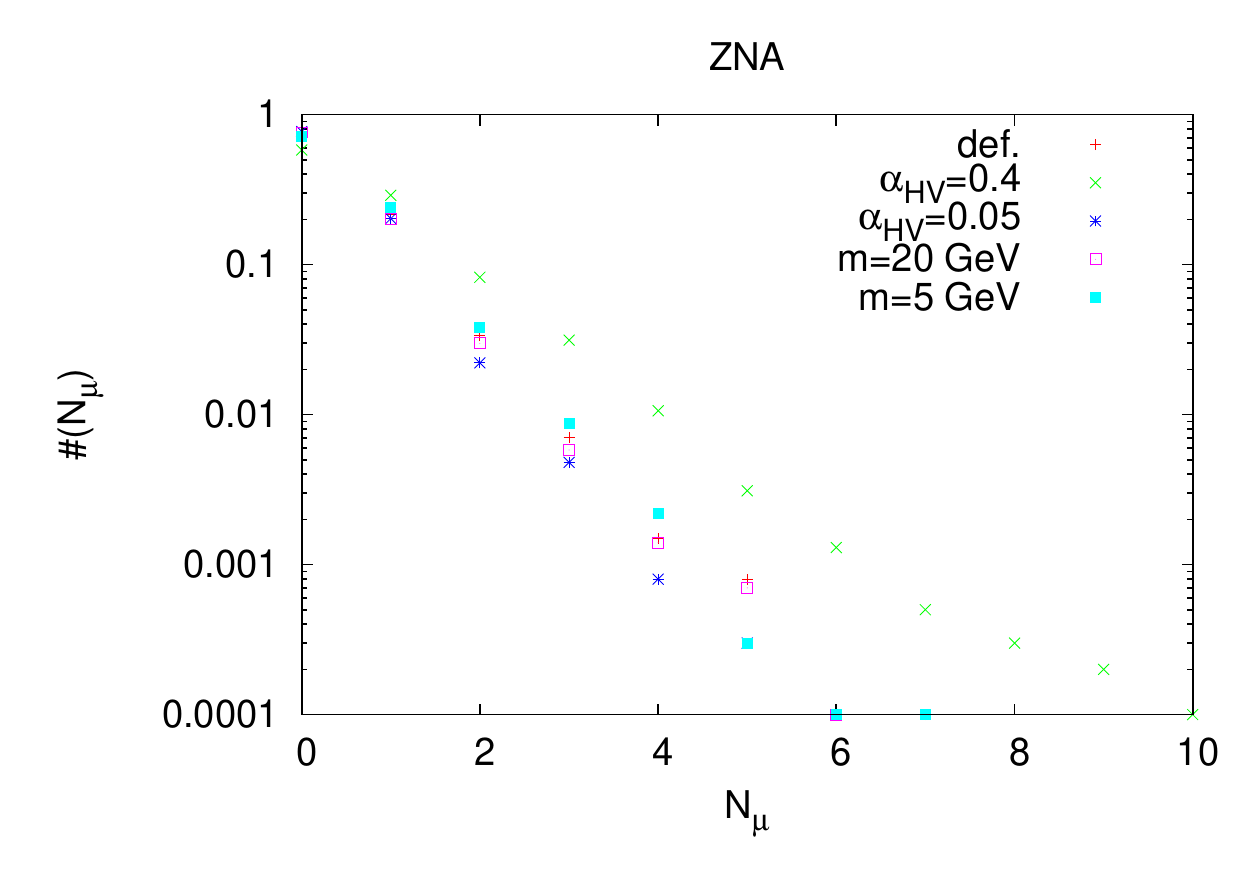}
\includegraphics[width=78 mm, angle = 0]{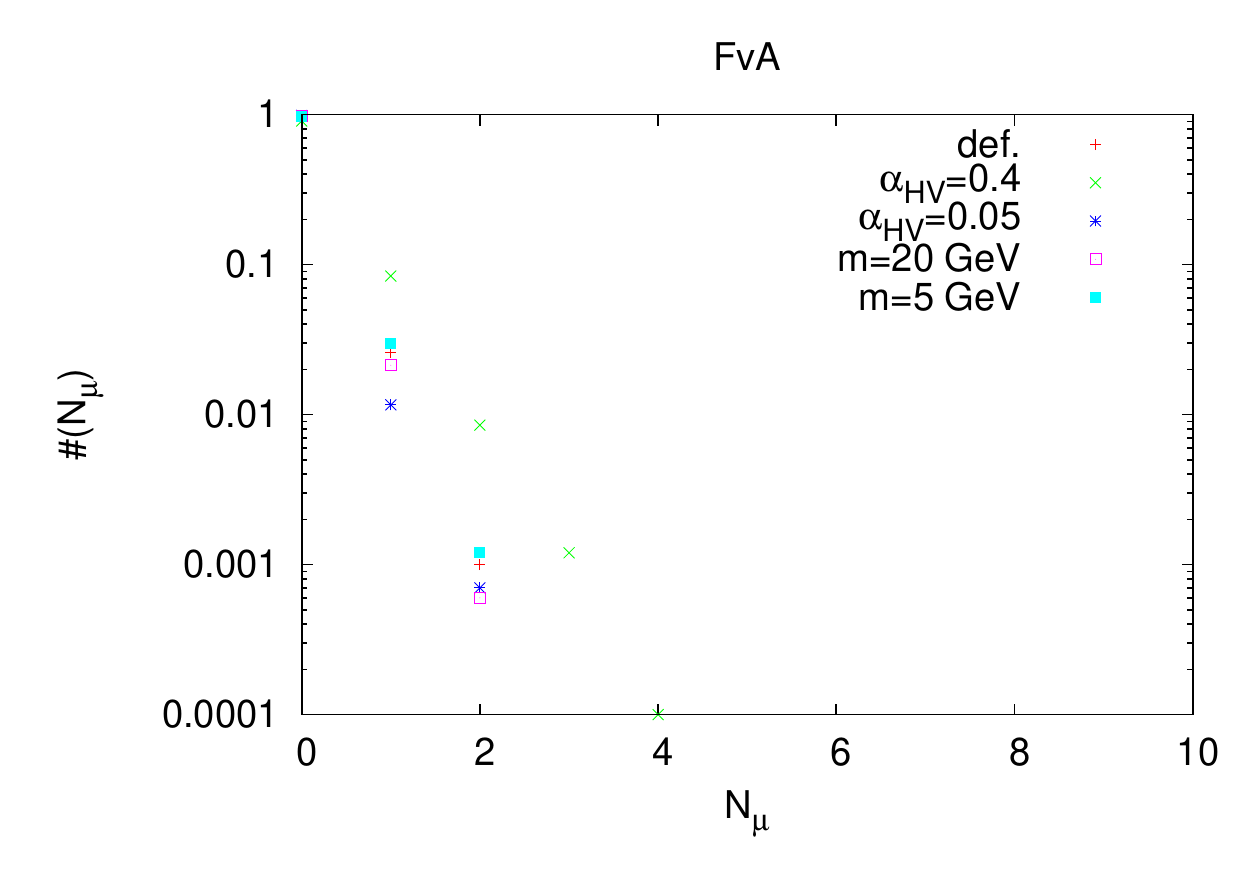}
\includegraphics[width=78 mm, angle = 0]{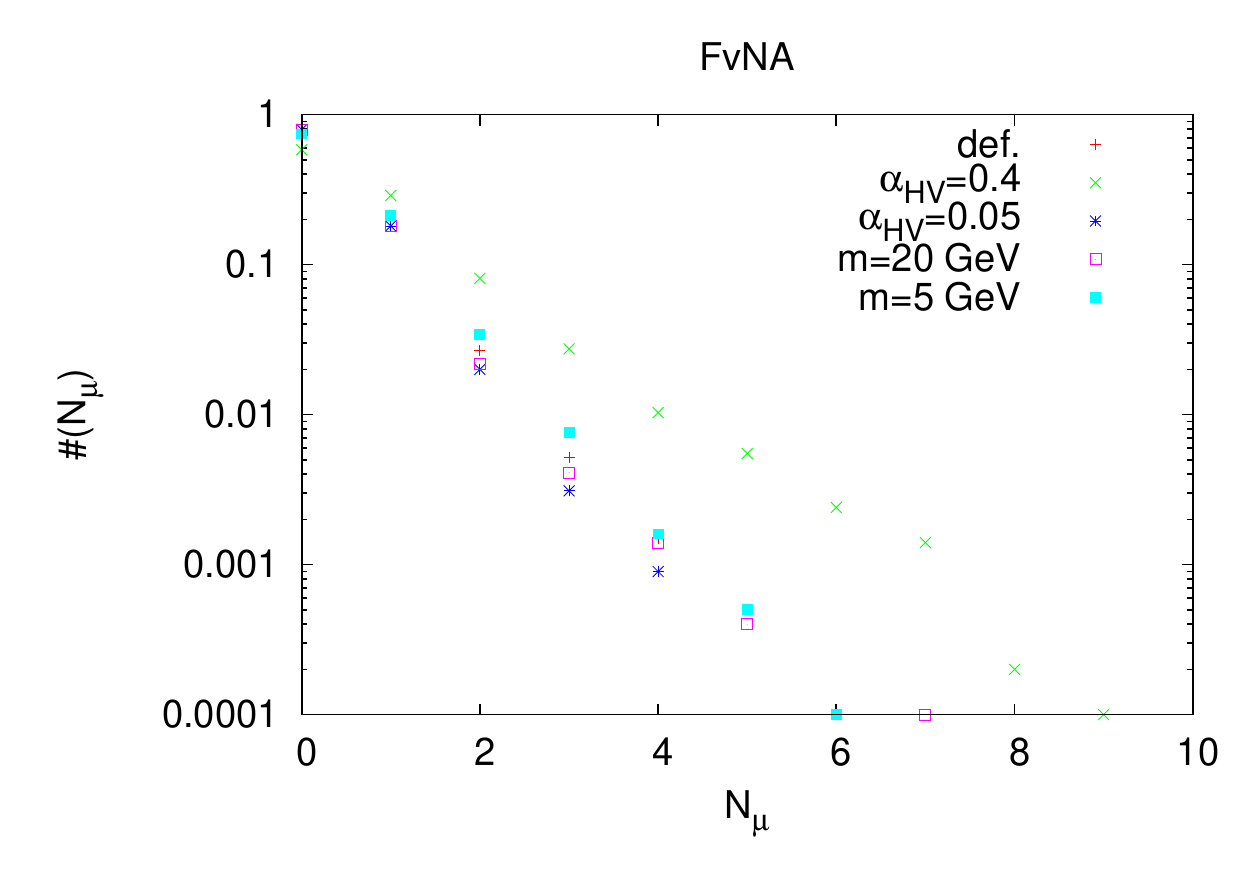}

\caption{Number of muons pairs with invariant mass close to $ m_{\gammav /\piv /\rhov} $. To the top ZA (left) and ZNA (right) and below FvA (left) FvNA (right),
 the coupling strength and the $ \gammav /\piv /\rhov $ mass is varied as in fig \ref{nRad}.
The default values are $ \alpha_{HV}=0.1 $ and $ m_{\gammav /\piv /\rhov}=10 $ GeV}
\label{nLept}
\end{figure}

The coupling strength will determine the amount of radiation in the Abelian cases, as shown in Fig. \ref{nRad}. For the non-Abelian ones
the coupling still has an effect but the $ \qv $ will always hadronize as they need to be confined. The charged multiplicity
gives a general measure on the amount of activity in an event, but since a hadron collider produces much background the differences 
is not easily distinguishable, as seen in Fig. \ref{nCharged}. Here is shown all charged particles, which receives its major
 part from the background. One might instead try to select particles such as 
above some $ \pT $ threshold in order to remove background effects. Since this is fairly similar to jets, which will be studied below it has not been done. Lepton pairs
with proper invariant mass are on the other hand almost only from Valley particles, and are shown in Fig. \ref{nLept}.
 The Abelian distributions are directly proportional to the
respective HV particle content but in the non-Abelian case the decay channels to leptons differ for the $ \piv $ and $ \rhov $ so
the comparison is slightly more difficult. If the presence of leptons is necessary to single out HV events, then only multi-pair
 events will offer further information and such events may be rare.

\begin{figure}
\includegraphics[width=78 mm, angle = 0]{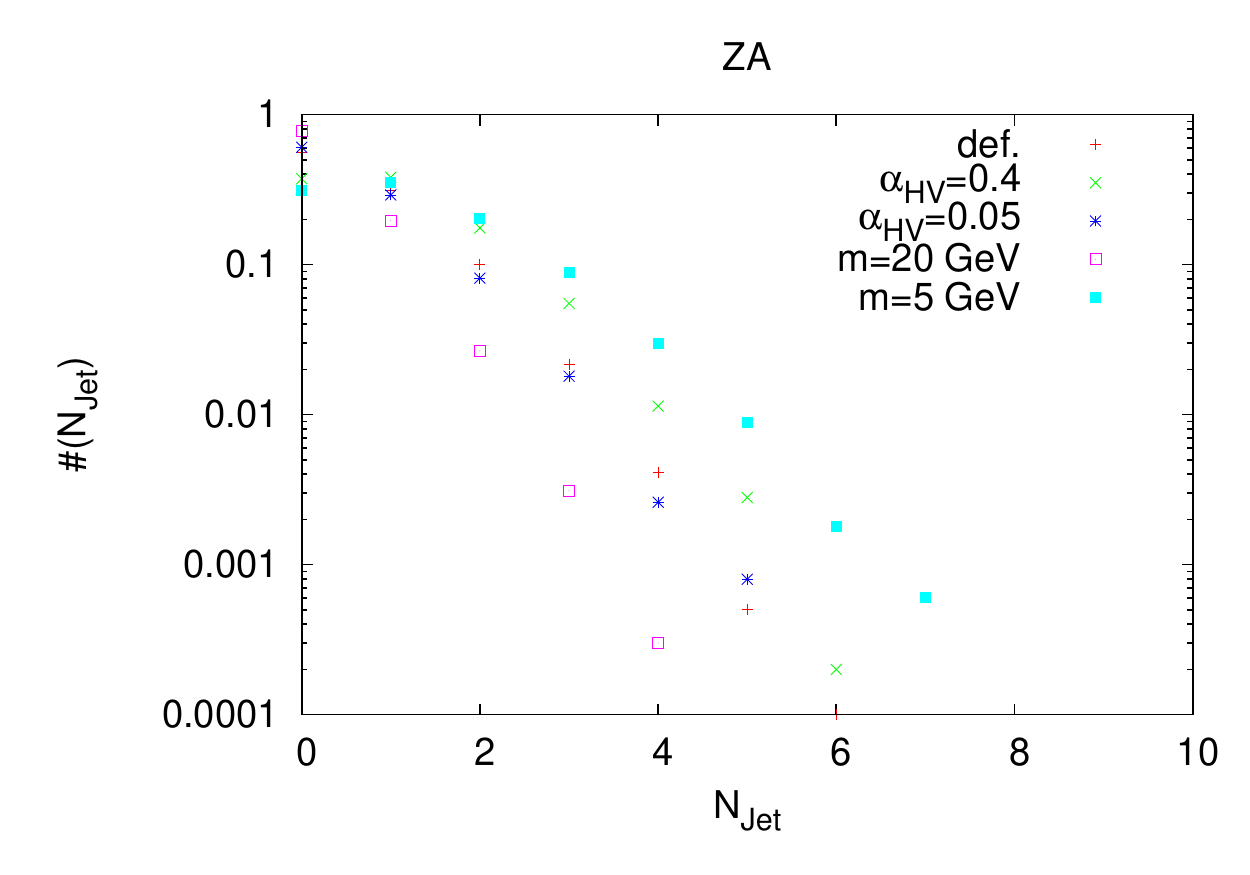}
\includegraphics[width=78 mm, angle = 0]{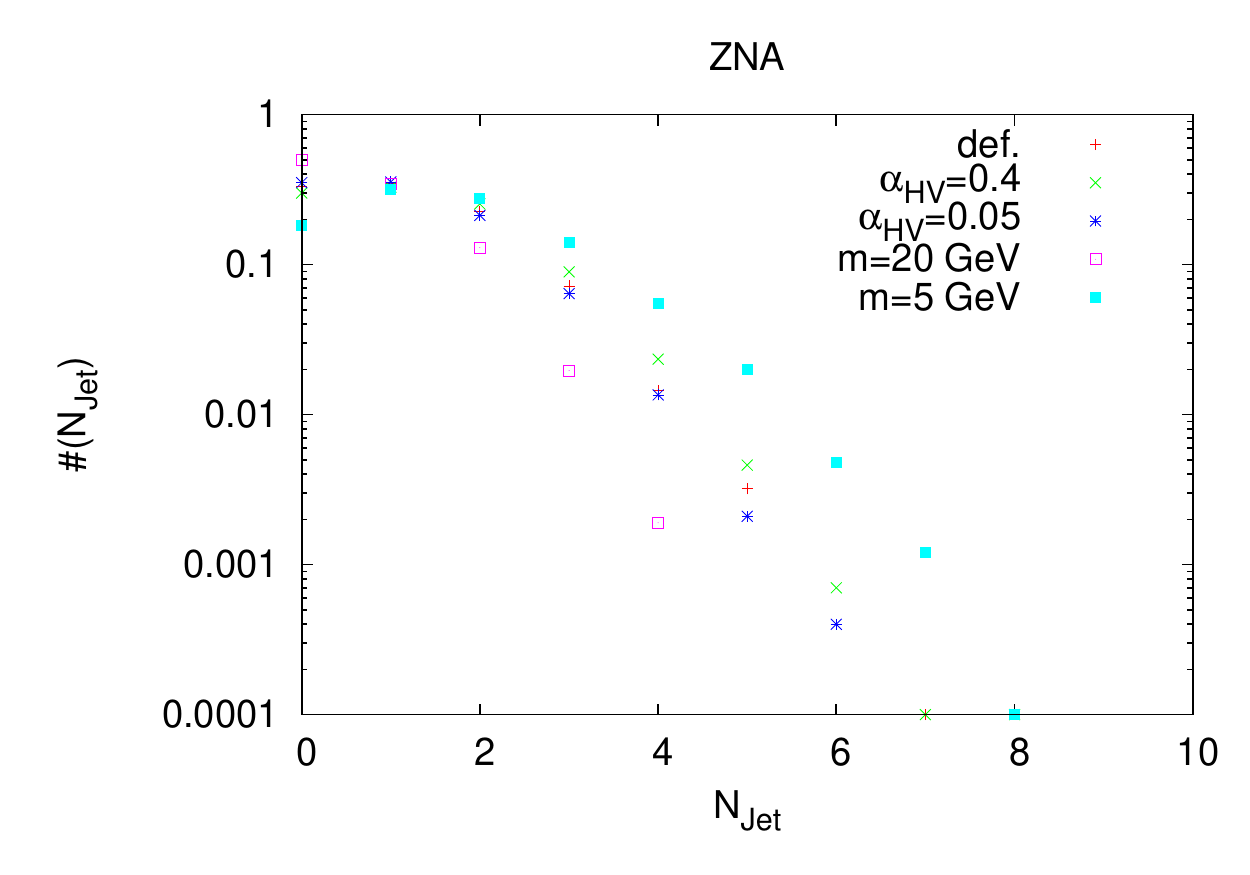}
\caption{The amount of jets in the different scenarios in the ZA (left) and ZNA (right),
 The coupling strength and the $ \gammav /\piv /\rhov $ mass is varied as in fig \ref{nRad}.
The default values are $ \alpha_{HV}=0.1 $ and $ m_{\gammav /\piv /\rhov}=10 $ GeV.
}
\label{nJet}
\end{figure}

For jets we use the radius of $ R=0.7 $ and put the $ \eT $ limit high enough, $ \eT=4m/R $, so that the products of a 2 particle decay from a 
particle with the $ \gammav/\piv/\rhov $ mass will be confined in one jet. There is usually more hadrons than that but it serves 
as approximation. The amount of jets present is shown in Fig. \ref{nJet}. Due to the need of changing $ \eT $ limits
 with changed mass, the mass makes a huge difference since it means more or less jets found. The $ \alpha_{HV} $ influence on the amount of
decaying HV particles is not seen, since there is a necessary $ \eT $ to be detected. Distributing the energy
 over more particles might actually reduce the amount of detected jets.

\begin{figure}
\centering
\includegraphics[width=135 mm, angle = 0]{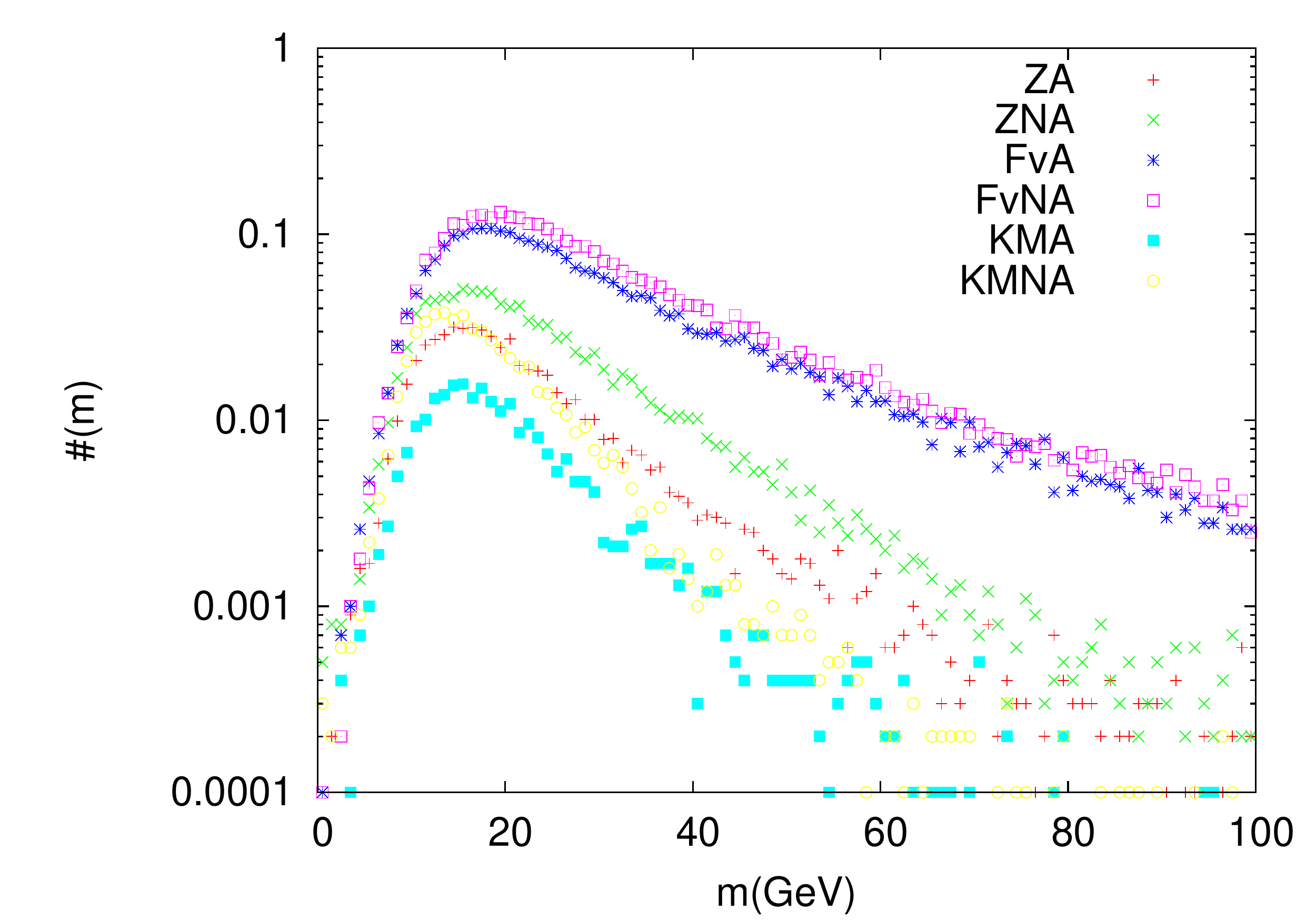}
\caption{The invariant mass of Jets with jet parameters with lower cut to find jets from a HV particle of mass 10 GeV.}
\label{mJet}
\end{figure}

The invariant mass of  jets can be calculated with results shown in Fig. \ref{mJet}. It once again peaks around the 
mass of the $ \gammav $ or $\piv/\rhov $, although this time it is far from the clean spike observed with leptons. The problem arises due to
all the background hadrons and possible overlap between jets.

\begin{figure}
\centering

\includegraphics[width=78 mm, angle = 0]{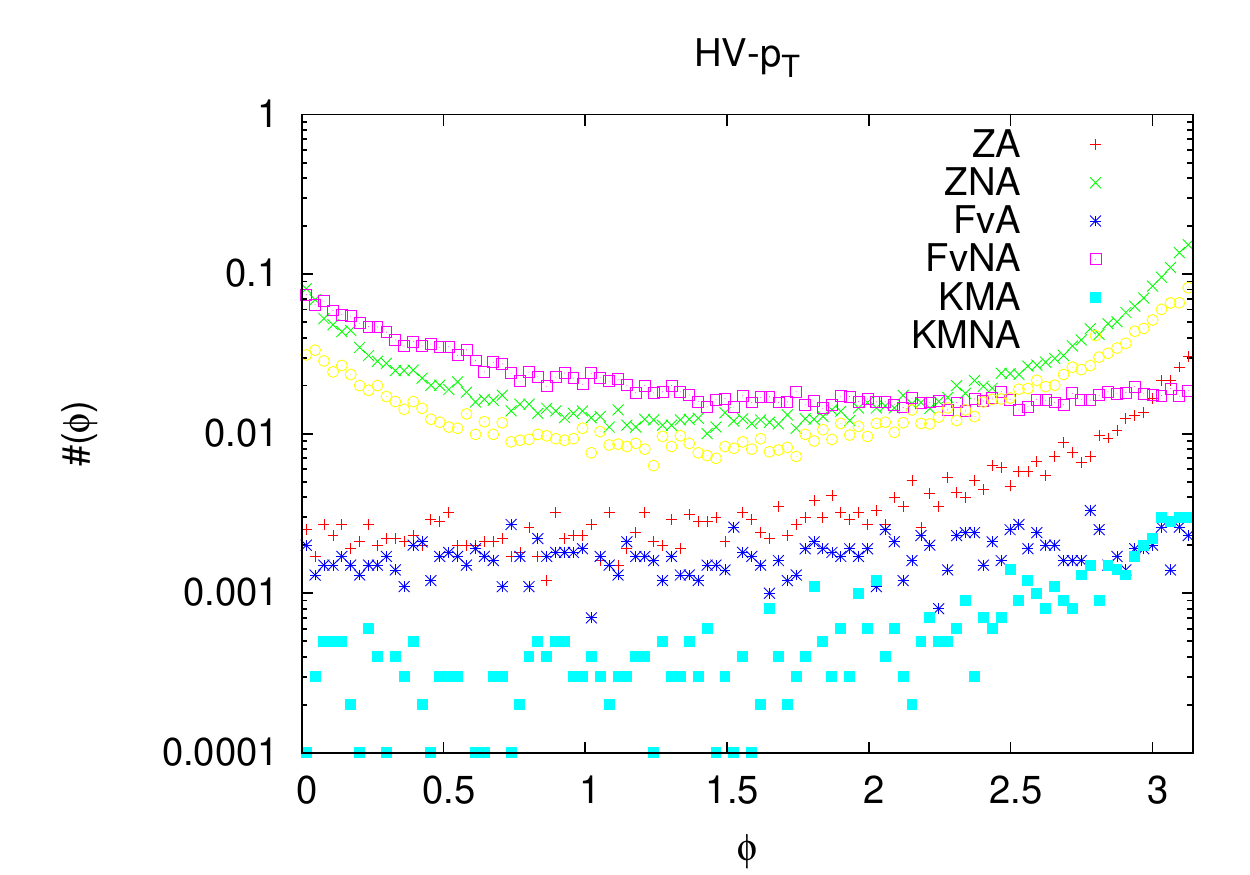}
\includegraphics[width=78 mm,angle = 0]{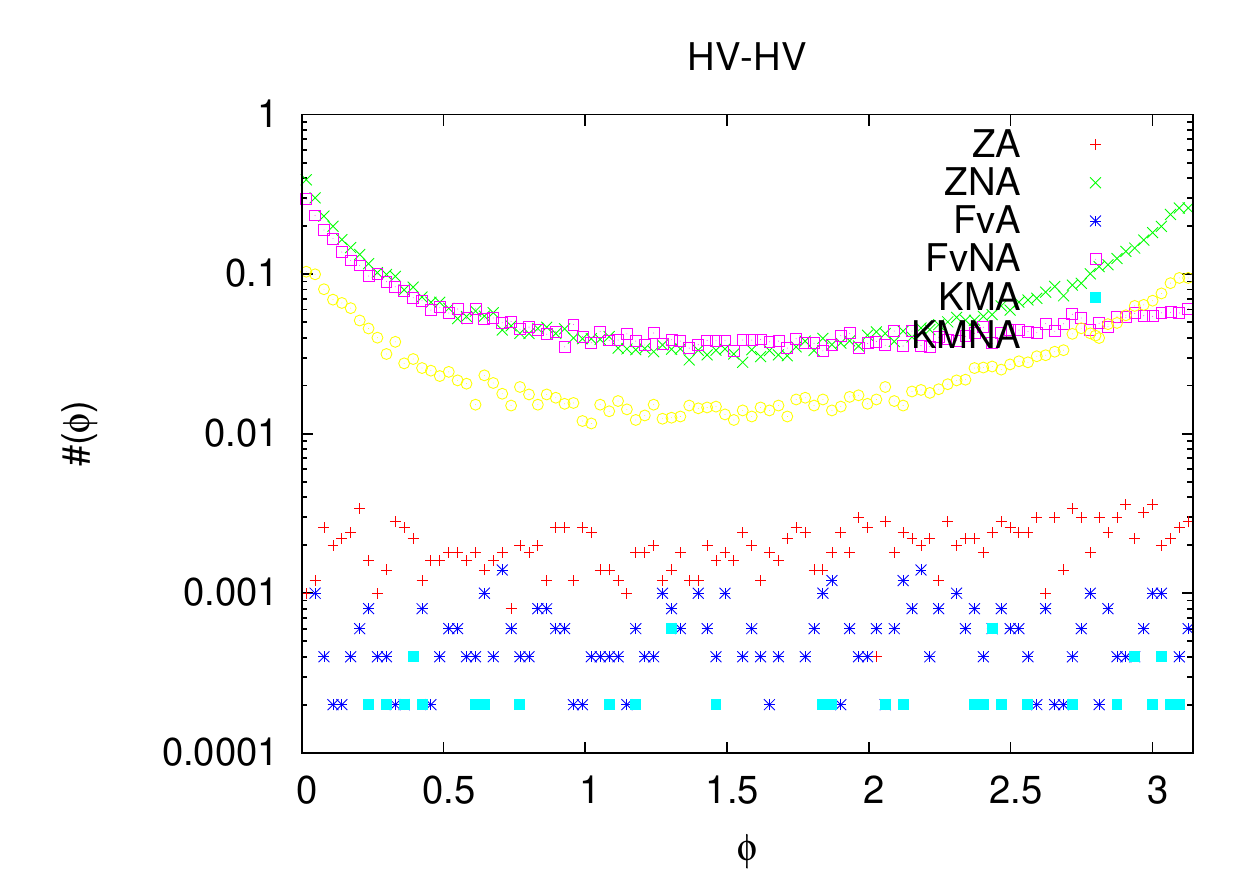}
\includegraphics[width=78 mm,angle = 0]{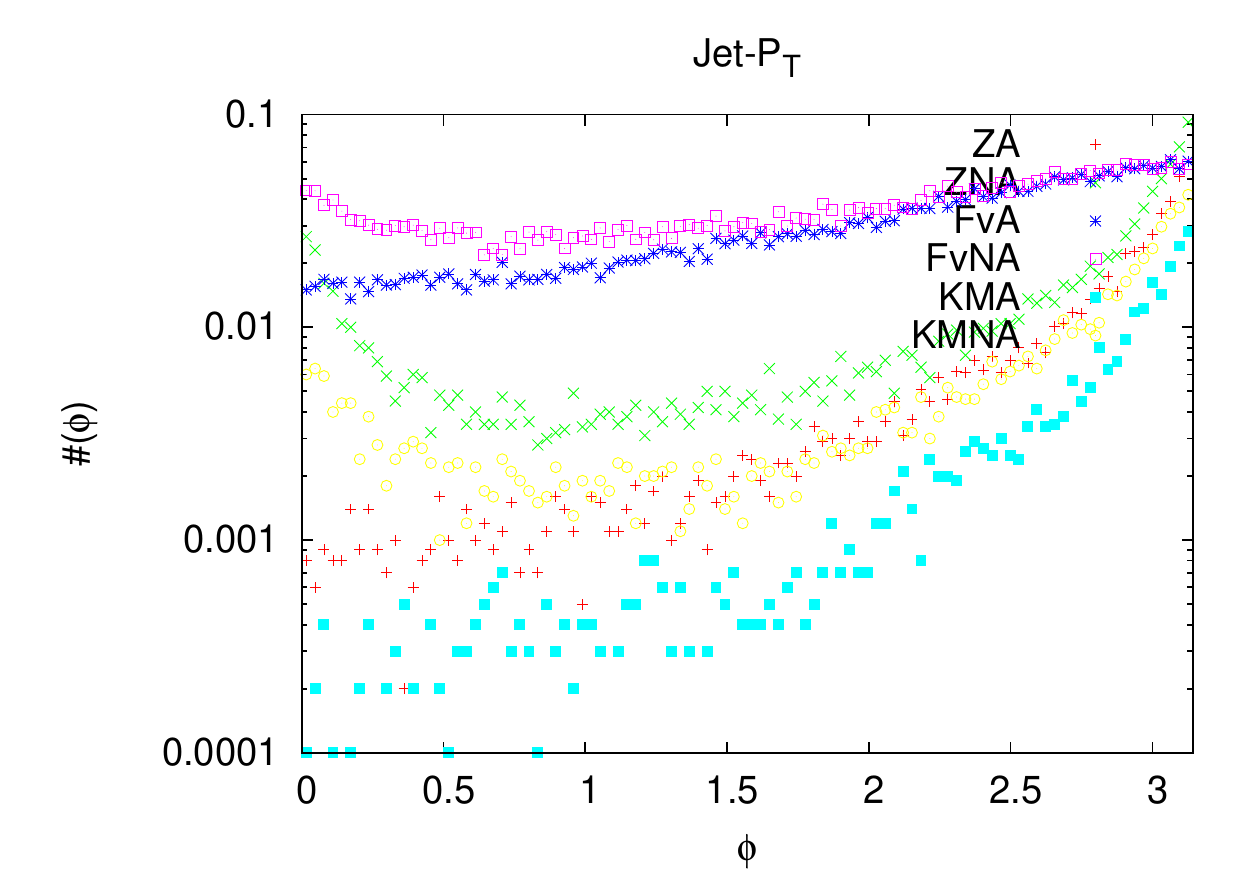}
\includegraphics[width=78 mm, angle = 0]{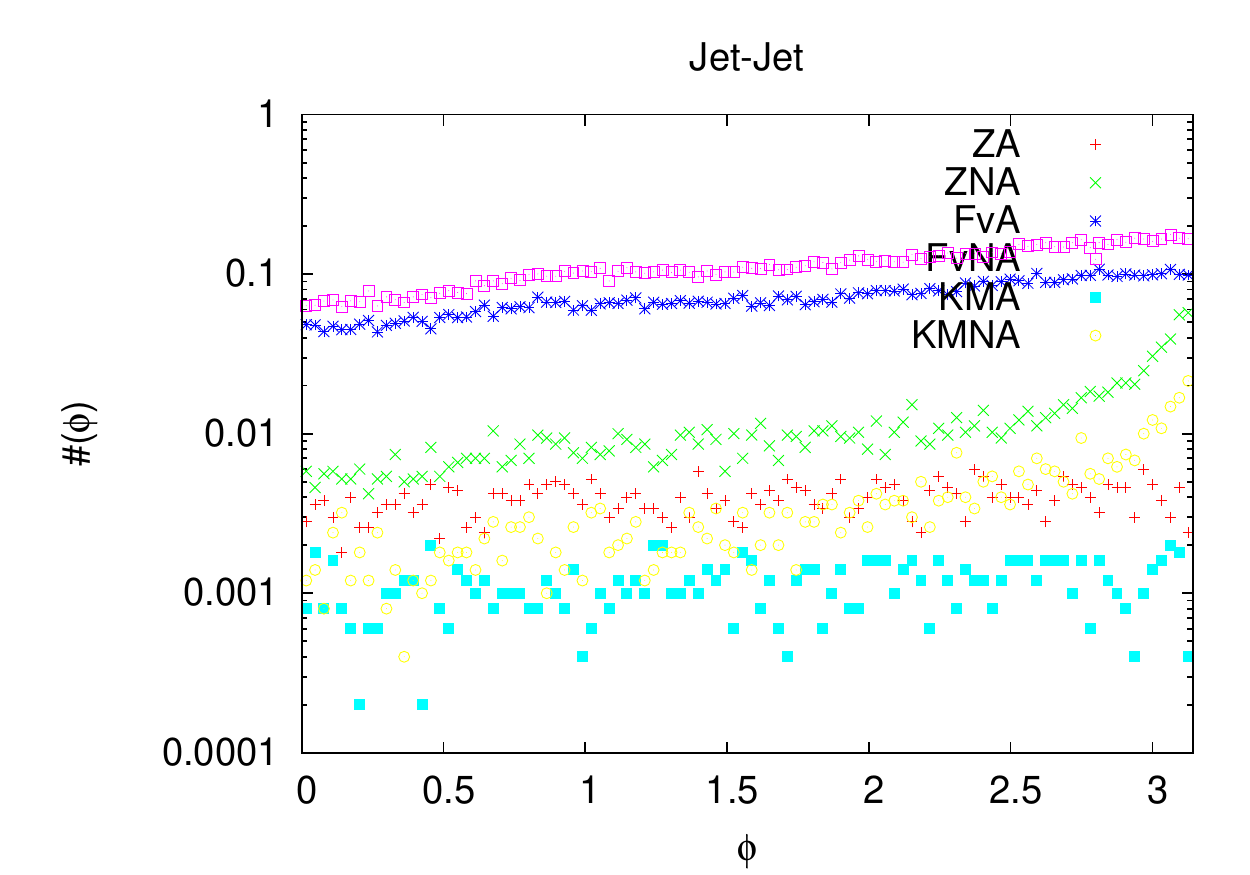}
\caption{To the top left is the azimuthal angle between the $ \pT $ direction and the $ \gammav/\piv/\rhov $.
To the top right is the relative azimuthal angle between the pairs of $ \gammav/\piv/\rhov $. Below is the corresponding 
angles with jets replacing the HV particles.}
\label{phiJet}
\end{figure}

The angle between  $ \pT $ and the HV particles and, since the latter are not directly detectable,
 the corresponding angle between $ \pT $ and jets is shown to the left in Fig. \ref{phiJet}.
In the ZNA and KMNA scenarios the $ \qv $ will be back-to-back and the jets of hadrons will roughly be in the $ \qv $ 
direction. The side with least diagonal mesons will then usually be the $ \pT $ direction, so a jet will be present in the opposite direction.
In the other direction there might be less both in momenta (leads to jet-finders missing them) and actual number of jets.
 For the FvNA, on the other hand, the $ \qv $ will provide the $ \pT $ in the  $ \Dv \rightarrow d\qv $ decay while the $ d $
quark appears as a highly energetic jet in the opposite direction in the  $ \Dv $ rest frame. The effects are in general not as clear since there are
 two $ \Dv $s and the different directions will give events with no observed match.
 In the Abelian case the emissions of $ \gammav $ do not favor the $ \qv $ direction. If few $ \gammav $s are emitted per event the $ \pT $ 
direction will be opposite to the most energetic $ \gammav $, but now there is no mechanism that favors $ \gammav $ in the opposite direction.
 As such it can be used to differentiate an Abelian and non-Abelian scenario.

The angle between the jet pairs are shown to the right in Fig. \ref{phiJet}. Here the same effect is present as for the comparison to the $ \pT $,
 as the jets are frequently in opposite direction for KMNA and ZNA, and fairly evenly distributed for the Abelian.
Since there are few jets energetic enough to be captured in our jet finder the same direction mesons are not visible in the jets.
The background will also be larger for many jets, due to there being $ n(n-1)/2 $ jet pairs with $ n $ jets present, which hides
more of the effects.

\begin{figure}
\centering
\includegraphics[width=78 mm, angle = 0]{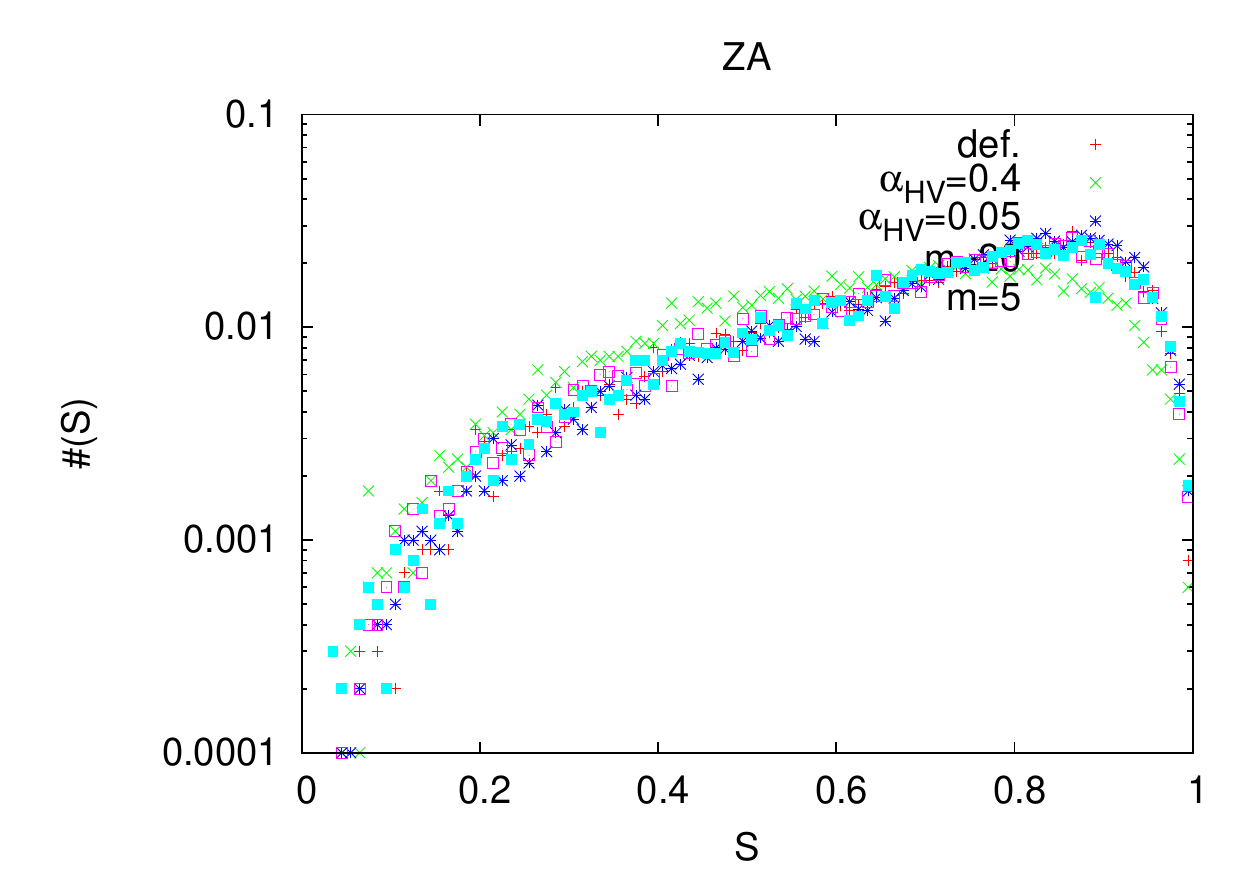}
\includegraphics[width=78 mm, angle = 0]{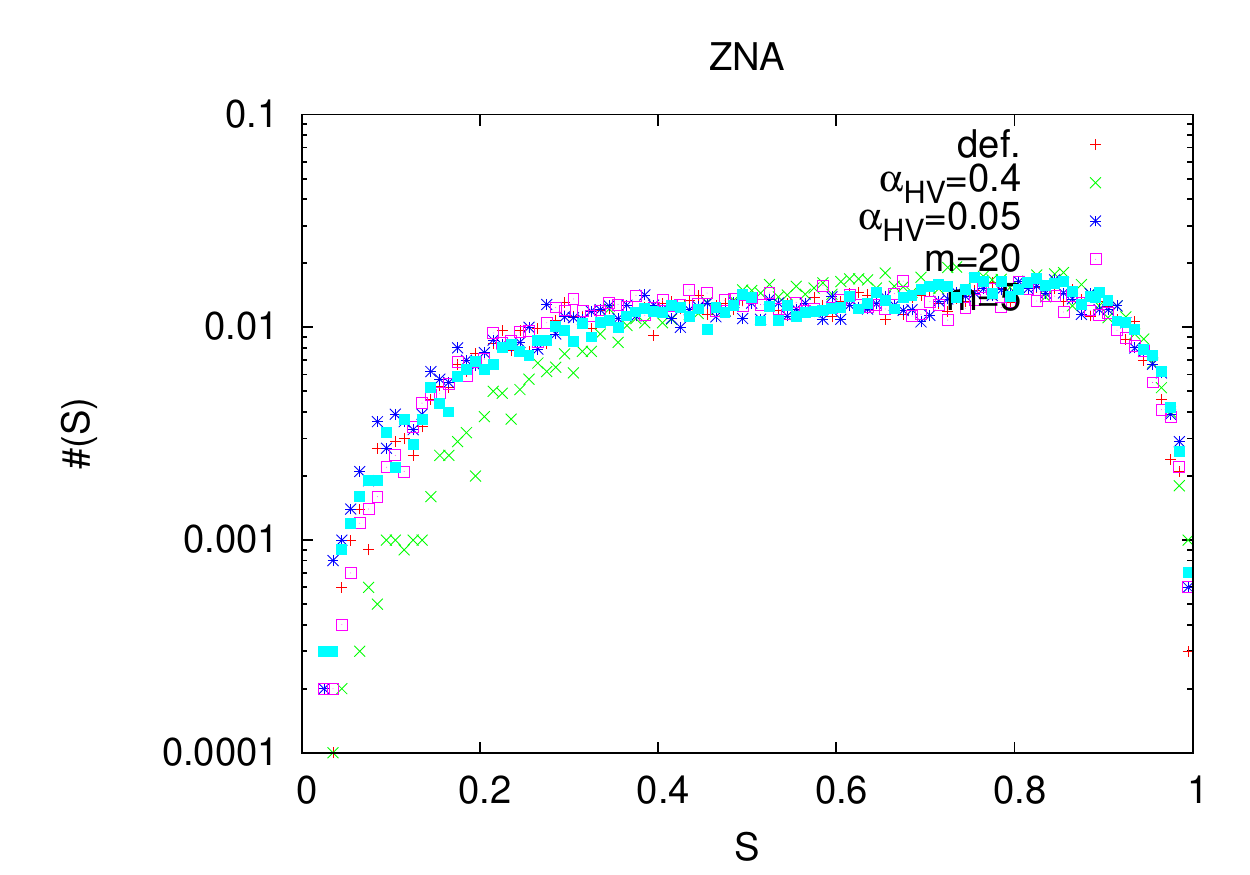}
\includegraphics[width=78 mm, angle = 0]{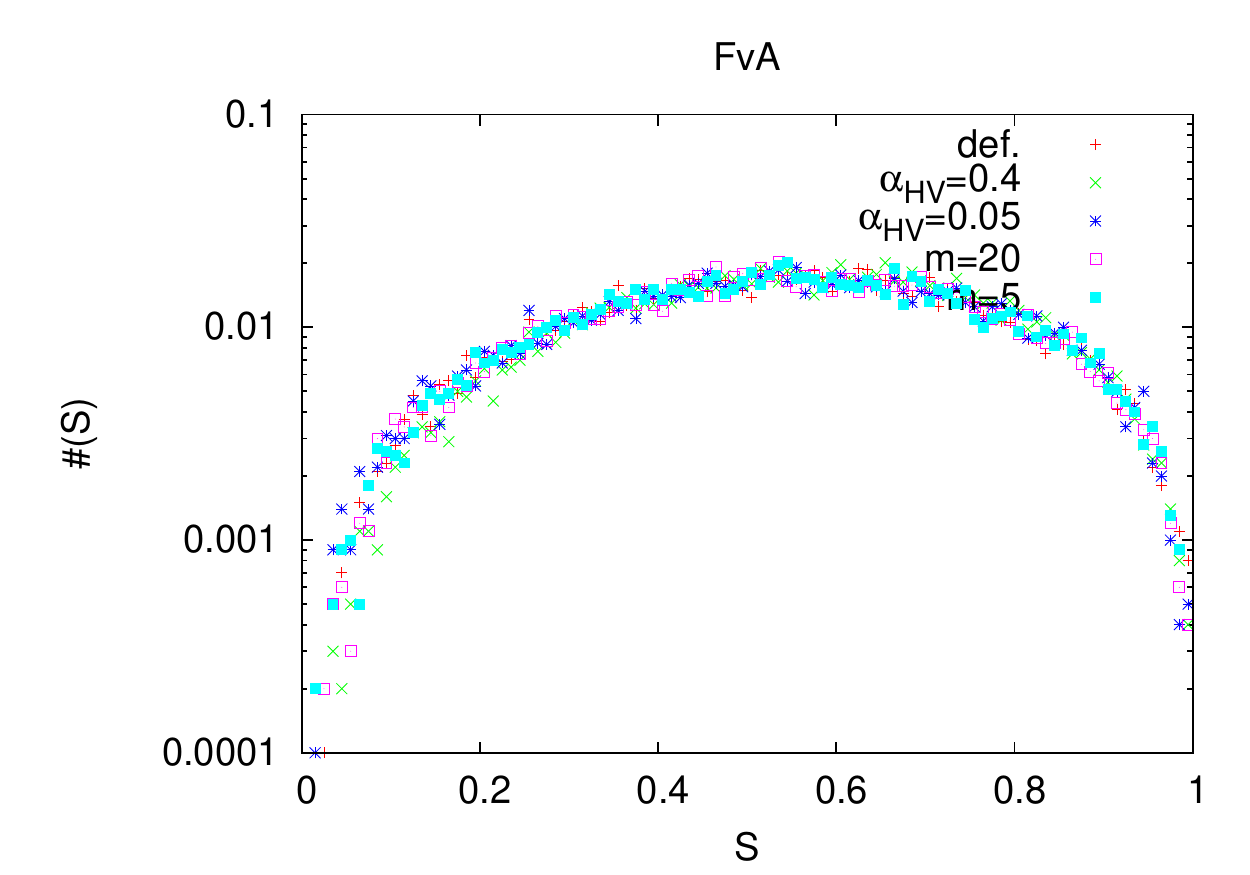}
\includegraphics[width=78 mm, angle = 0]{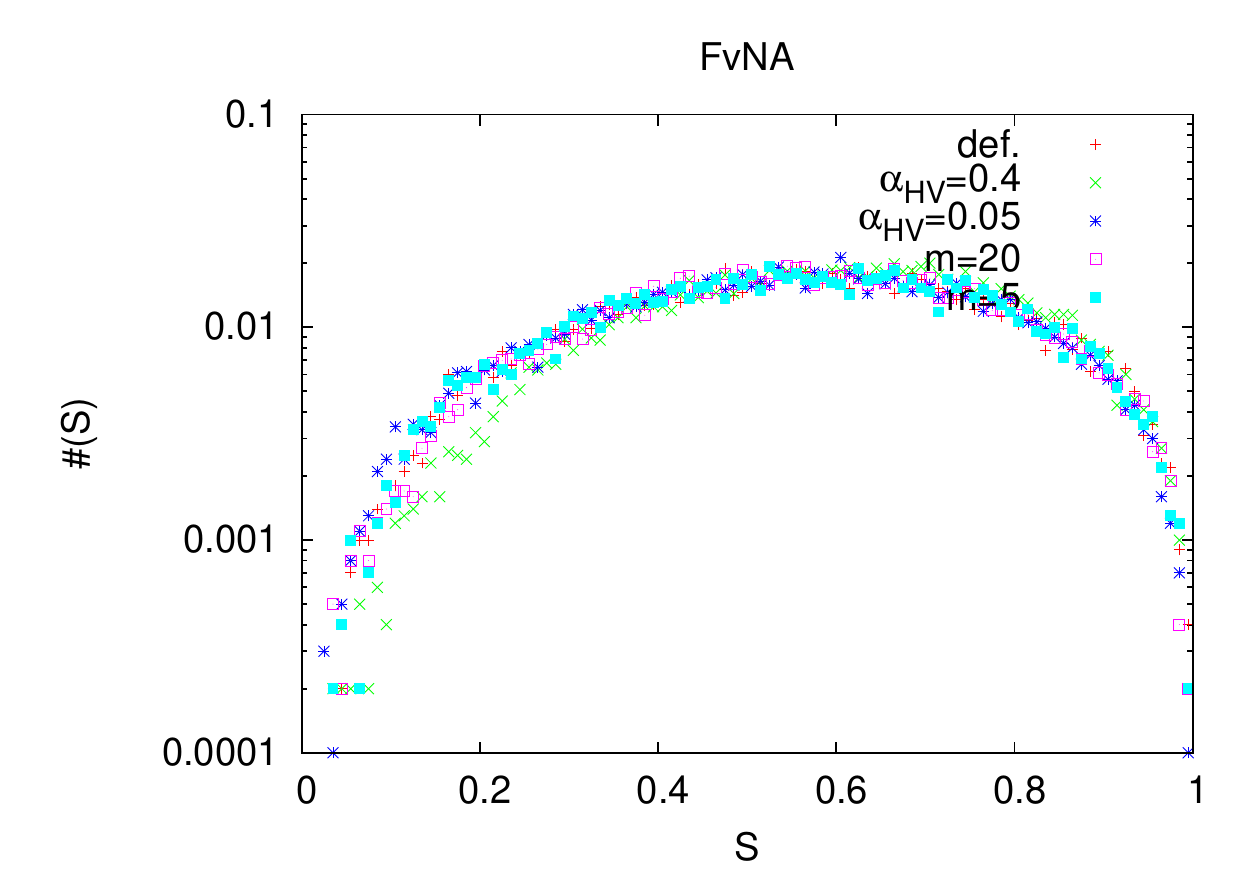}
\includegraphics[width=78 mm, angle = 0]{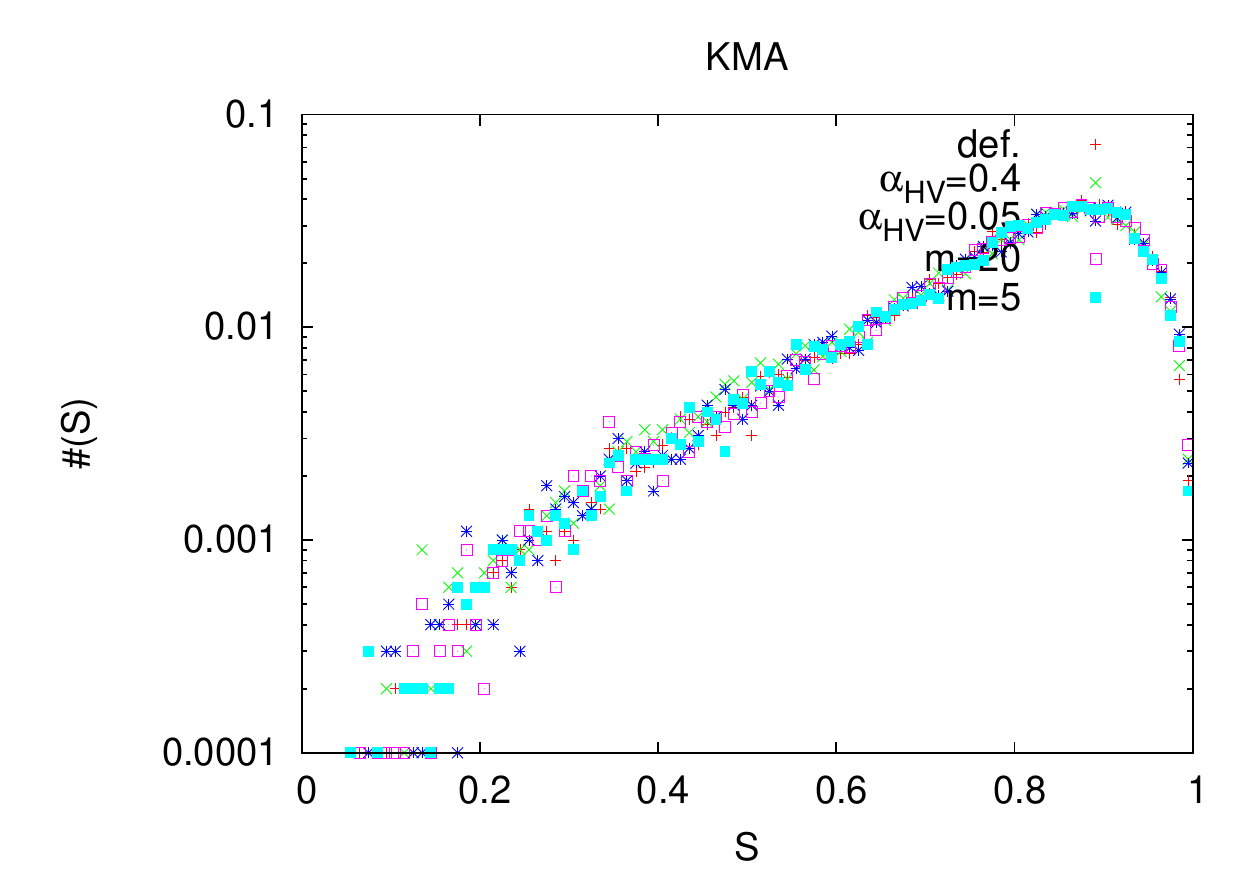}
\includegraphics[width=78 mm, angle = 0]{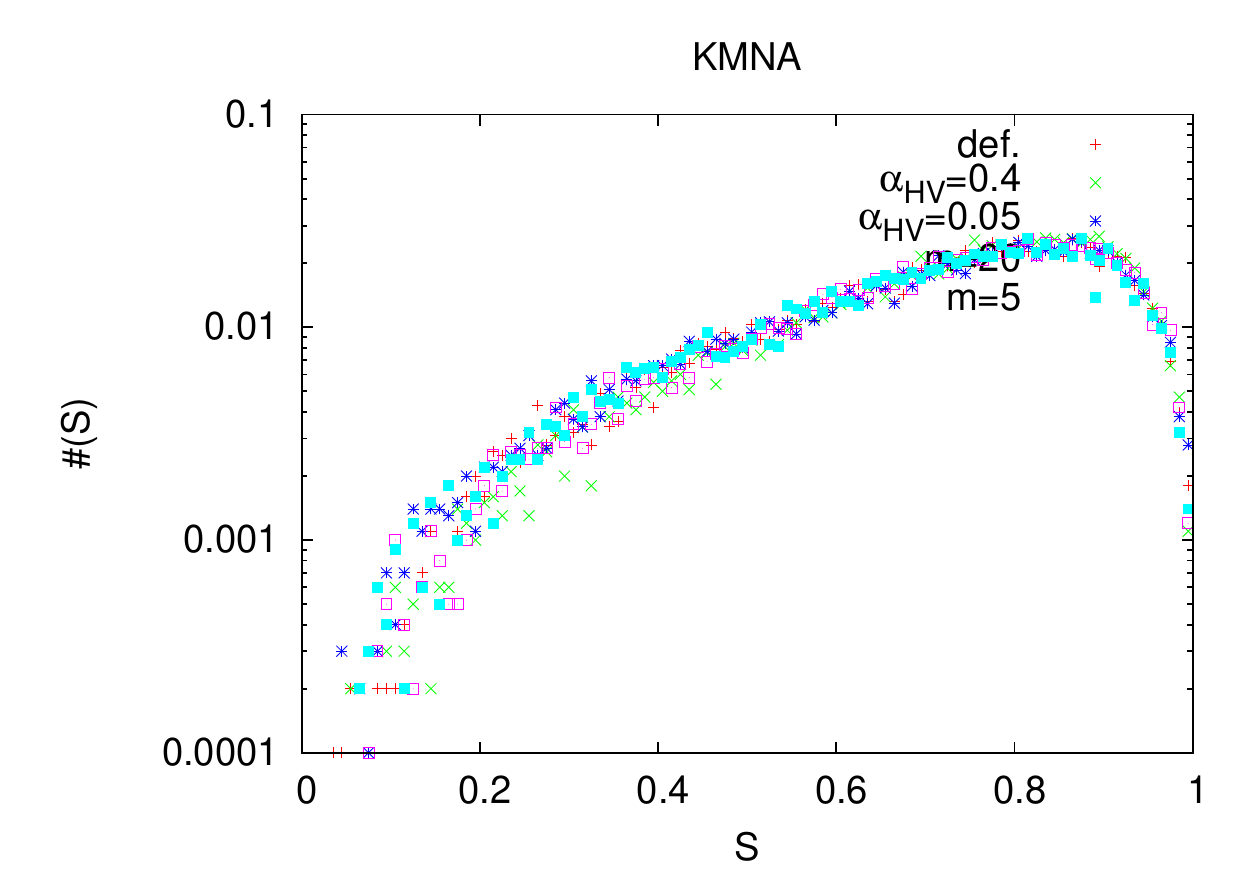}
\caption{The linearized sphericity of the six models, abelian to the left and non-Abelian to the right. 
From top to botton is the Z, Fv and KM scenarios. Different values for $ \alpha $ or the $ \gammav/\piv/\rhov $ masses are shown.}
\label{sphericity}
\end{figure}

The linearized sphericity is shown in Fig. \ref{sphericity}.
The Fv events are as expected more round as the $ \Fv \rightarrow f\qv $ decays don't give back-to-back events. Also the KM are 
more spherical than the Z due to the larger energies in the Z case and thus the background will be less noticeable. In the non-Abelian case a higher coupling constant leads to
more spherical event. One would expect the same in the Abelian case but as seen for the ZA the events become less spherical. 
This might be caused by so low emission rates so that the background effects, which tend to be round, dominates. We have not yet investigated this further.
Since emission of
particles also depends on masses one might expect some effect by changing the $ \gammav/\piv/\rhov $ mass, but it appears
to make no large differences.

\section{Summary and conclusions}
In this study we have investigated some Hidden Valley models and possible measurements at LHC with full energy of 14 TeV.
The six models in our study had either an $ U(1) $ or $ SU(N) $ hidden gauge group and, for each of them, production through $ \Fv $,  $ Z' $
and kinetic mixing was considered. For the study we used a previously implemented model in \textsc{pythia}

The Fv scenario was easily distinguishable since in most plots, differences was present due to the $ \Fv \rightarrow f\qv $ decay. The Z and KM
were fairly similar, but the presence or absence of a mass spike at $ m_{Z'} $ could be used. The difference
between the Abelian and non-Abelian models turned out to be trickier. There were a few effects, like the slight difference in $ \pT $ distribution
and the lower limits of $ \qv $ mass pair. The trouble is that both require a large amount of statistics, and in
the latter case also low-$ \pT $ measurements that might be hard to separate from the background. In the angular distributions of jets
relative to the $ \pT $ 
there was a significant difference between the Abelian and non-Abelian models due to differences in the angular distributions of
$ \gammav $ emissions compared to the hadronization into $ \qv $. 

The impact of different parameter sets was not investigated, and it is not unreasonable to assume that discerning the scenarios
at least gets more difficult if not actually impossible. The Fv pair and mass spikes for the Z' will still be present independent of parameters.
 Likewise the differences between Abelian and non-Abelian, the differences in angular distribution
for Valley particles, should not be so parameter-dependent, although visible results might.

We also looked at some means to measure the different parameters. 
The masses for particles that decayed to SM were easily detectable trough lepton pairs. The masses of several other particles could be determined
from the invariant mass distribution, which could be constrained from the $ \pT $ distribution, although a lot of events 
will be necessary to do so. The coupling strength turned out to be more difficult to access, it gave a clear effect on the amount of valley particles
but the visible effects was not as clear due to background effects or low amount of events in the lepton case. 

Since many of the effects required many measurements and that HV events are rare to begin with the obvious next step is to 
investigate actual production cross sections to see whether is possible to gather sufficient events.
In this case one also needs to consider the background of SM events,
 since one only can work with the events that can be identified as HV ones. 
Also different parameter values will at least make a difference in how many events are needed for the different methods of
distinguish models and measure parameters. This also requires study of the experimental errors that could be expected.
Finally the model has to be handed to the experimental community in  order to check with the LHC data in order to be confirmed or denied.

\newpage

{\Large \bf Acknowledgements} \linebreak
I would like to then my supervisor Torbjörn Sjöstrand for all his help and for explaining and answering all my questions really
well. I also want to thank Jacob Winding for helping me with some computer-related stuff for keeping me company during my work.

\end{document}